\documentclass[]{aa}
\pdfoutput=1
\usepackage{geometry} 
\usepackage{graphicx}
\usepackage[varg]{txfonts}
\usepackage{upgreek}
\usepackage{soul}
\usepackage{placeins}
\bibpunct{(}{)}{;}{a}{}{,} 
\usepackage{lscape}

\newcommand{\radu}{rad m$^{-2}$}

\newcommand{\inv}{$^{-1}$}

\usepackage{microtype}
\title{Polarized point sources in the LOFAR Two-meter Sky Survey: A preliminary catalog}
\titlerunning{LOTSS polarized point sources}
\author{C.L.~Van Eck\inst{1} \and 
M.~Haverkorn\inst{1}  \and
M.I.R.~Alves\inst{1} \and
R.~Beck\inst{2} \and
P.~Best\inst{3} \and
E.~Carretti\inst{4} \and
K.T.~Chy\.zy\inst{5} \and
J.S.~Farnes\inst{1} \and
K.~Ferri\`{e}re\inst{6} \and
M.J.~Hardcastle\inst{7} \and
G.~Heald\inst{8,9} \and
C.~Horellou\inst{10} \and
M.~Iacobelli\inst{11} \and
V.~Jeli\'{c}\inst{12} \and
D.D.~Mulcahy\inst{13} \and
S.P.~O'Sullivan\inst{14} \and
I.M.~Polderman\inst{1} \and
W.~Reich\inst{2} \and
C.J.~Riseley\inst{8} \and
H.~R\"{o}ttgering\inst{15} \and
D.H.F.M.~Schnitzeler\inst{2,16} \and
T.W.~Shimwell\inst{11,15} \and
V.~Vacca\inst{4} \and
J.~Vink\inst{17} \and
G.J.~White\inst{18,19}}
\authorrunning{C.L. Van Eck. et al.}
\institute{Department of Astrophysics/IMAPP, Radboud University, PO Box 9010, NL-6500 GL Nijmegen, the Netherlands; 
\email{c.vaneck@astro.ru.nl} 
\and
Max-Planck-Institut f\"{u}r Radioastronomie, Auf dem H\"{u}gel 69, 53121 Bonn, Germany \and 
SUPA, Institute for Astronomy, Royal Observatory, Blackford Hill, Edinburgh, EH9 3HJ, UK \and 
INAF-Osservatorio Astronomico di Cagliari, Via della Scienza 5, I-09047 Selargius (CA), Italy \and 
Astronomical Observatory, Jagiellonian University, ul. Orla 171, 30-244 Krak\'ow, Poland \and 
IRAP, Universit\'{e} de Toulouse, CNRS, 9 avenue du Colonel Roche, BP 44346, 31028, Toulouse Cedex 4, France \and 
Centre for Astrophysics Research, School of Physics, Astronomy and Mathematics, University of Hertfordshire, College Lane, Hatfield AL10 9AB, UK \and 
CSIRO Astronomy and Space Science, 26 Dick Perry Ave, Kensington, WA 6151 Australia \and
University of Groningen, Kapteyn Astronomical Institute, NL-9700 AV Groningen, Netherlands \and
Dept of Space, Earth and Environment, Chalmers University of Technology, Onsala Space Observatory,  439 92, Onsala, Sweden \and 
ASTRON, the Netherlands Institute for Radio Astronomy, Postbus 2, 7990 AA Dwingeloo, The Netherlands \and 
Ru{\dj}er Bo\v{s}kovi\'{c} Institute, Bijeni\v{c}ka cesta 54, 10000 Zagreb, Croatia \and 
Jodrell Bank Centre for Astrophysics, Alan Turing Building, School of Physics and Astronomy, The University of Manchester, Oxford Road, Manchester, M13 9PL, UK \and 
Hamburger Sternwarte, Universit\"{a}t Hamburg, Gojenbergsweg 112, 21029, Hamburg, Germany \and 
Leiden Observatory, Leiden University, PO Box 9513, 2300 RA Leiden, The Netherlands \and 
Bendenweg 51, 53121 Bonn, Germany \and
Astronomical Institute Anton Pannekoek, University of Amsterdam, Science Park 904, 1098XH Amsterdam, The Netherlands \and 
RAL Space, Rutherford Appleton Laboratory, Chilton, Didcot, Oxfordshire, OX11 0QX, UK \and
Department of Physics and Astronomy, The Open University, Walton Hall, Milton Keynes, MK7 6AA, UK 
}

\date{} 

\abstract{The polarization properties of radio sources at very low frequencies (<200 MHz) have not been widely measured, but the new generation of low-frequency radio telescopes, including the Low Frequency Array (LOFAR: a Square Kilometre Array Low pathfinder), now gives us the opportunity to investigate these properties.
In this paper, we report on the preliminary development of a data reduction pipeline to carry out polarization processing and Faraday tomography for data from the LOFAR Two-meter Sky Survey (LOTSS) and present the results of this pipeline from the LOTSS preliminary data release region (10$^\mathrm{h}$45$^\mathrm{m}$ -- 15$^\mathrm{h}$30$^\mathrm{m}$ right ascension, 45\degr -- 57\degr\ declination, 570 square degrees). We have produced a catalog of 92 polarized radio sources at 150 MHz at 4\farcm3 resolution and 1 mJy rms sensitivity, which is the largest catalog of polarized sources at such low frequencies. We estimate a lower limit to the polarized source surface density at 150 MHz, with our resolution and sensitivity, of 1 source per 6.2 square degrees. We find that our Faraday depth measurements are in agreement with previous measurements and have significantly smaller errors. Most of our sources show significant depolarization compared to 1.4 GHz, but there is a small population of sources with low depolarization indicating that their polarized emission is highly localized in Faraday depth. We predict that an extension of this work to the full LOTSS data would detect at least 3400 polarized sources using the same methods, and probably considerably more with improved data processing.}


\begin{document}
\maketitle

\section{Introduction}\label{sec:intro}

Magnetic fields play a significant role in the dynamics and evolution of the interstellar medium (ISM) in galaxies, through many phenomena including acceleration and confinement of cosmic rays \citep[e.g.][]{Blasi2013}, angular momentum transport in star formation \citep[e.g.][]{Lewis17}, and magnetohydrodynamic turbulence \citep[e.g.][]{Falceta2014}.

Observing the linear polarization of radio sources provides insight into the magnetic fields both at the source (from synchrotron polarization and Faraday rotation) and along the line of sight between a source and the observer (from Faraday rotation).

For many years, catalogs of large numbers of polarized radio sources (with corresponding Faraday rotation measurements) have been useful for diverse purposes including modelling the large-scale structure of the magnetic field in the Milky Way \citep[e.g.][]{VanEck11, Jansson2012}, studying the properties of smaller-scale structure in the magnetic field of the Milky Way \citep[e.g.][]{Haverkorn08,Stil14}, studying magnetic fields in nearby galaxies \citep[e.g.][]{Han1998}, and studying the evolution over time of galactic magnetic fields \citep[e.g.][]{Arshakian2009}. A high sky surface density of such measurements is needed for the detection of magnetic fields in the intergalactic medium \citep{Vacca2016}.

The amount of Faraday rotation (i.e. the extent to which the polarization position angle has rotated between the emission source and the observer) is the product of the observing wavelength squared ($\lambda^2$) and the Faraday depth of the source ($\phi$) which is defined as 
\begin{equation}
\phi = 0.812\; {\rm rad \, m^{-2}} \int_{\mathrm{source}}^{\mathrm{observer}} \left( \frac{n_\mathrm{e}}{{\rm cm^{-3}}}\right) \left( \frac{\vec{B}}{{\rm \upmu G}} \right) \cdot \left(\frac{\vec{dl}}{{\rm pc}} \right),
\end{equation}
where $n_\mathrm{e}$ is the number density of free electrons, $\vec{B}$ is the magnetic field, $\vec{dl}$ is a differential element of the radiation path, and the integral is taken over the line of sight from the emission source to the observer.
\footnote{In the literature, the terms Faraday depth and Faraday rotation measure (RM) are sometimes used interchangeably, and sometimes used to distinguish between a physical quantity (the integral in Eq.~1) and a measurement (the rate of change of polarization angle with respect to wavelength squared). For this paper, we use Faraday depth for Eq.~1, and RM only when referring to literature that used that term (specifically, the measurements of \citet{Taylor09}).}

The wavelength-squared dependence of Faraday rotation allows low frequency (long wavelength) observations to measure Faraday depth very accurately. The new generation of very low frequency radio telescopes, including the Low Frequency Array \citep[LOFAR, ][]{vanHaarlem2013} and the Murchison Widefield Array \citep[MWA, ][]{Tingay2013} have the potential to measure Faraday depths to an accuracy of 0.1 \radu\ or better, and can identify multiple polarized features separated by only a few \radu. However, sources where the polarized emission is distributed over a range in Faraday depth experience strong wavelength-dependent depolarization, which can limit which magnetic field and ISM configurations can be observed \citep{Burn66}.

In this paper, we present a new catalog of polarized radio sources observed at very low frequencies (150 MHz) with LOFAR, covering a 570 square degree area of the sky. This catalog is more than an order of magnitude larger than the previously largest sample of polarized sources at such a low frequency \citep{Mulcahy14}. In Sect.~\ref{sec:pipeline} we present the data reduction pipeline we used to generate the catalog. In Sect.~\ref{sec:catalog} we present our catalog and compare it with the higher frequency (1.4 GHz) rotation measure catalog of \citet{Taylor09}. In Sect.~\ref{sec:analysis} we present some interesting properties of our catalog. In Sect.~\ref{sec:discussion} we discuss the implications of our results on larger area LOFAR polarization surveys and steps for improving our source-identification pipeline. Finally, in Sect.~\ref{sec:summary} we summarize our work and conclude with an overview of science capability uniquely enabled by this catalog.

\section{Data processing and source extraction}\label{sec:pipeline}
\begin{figure*}[htbp]
   \centering
   \includegraphics[width=\linewidth]{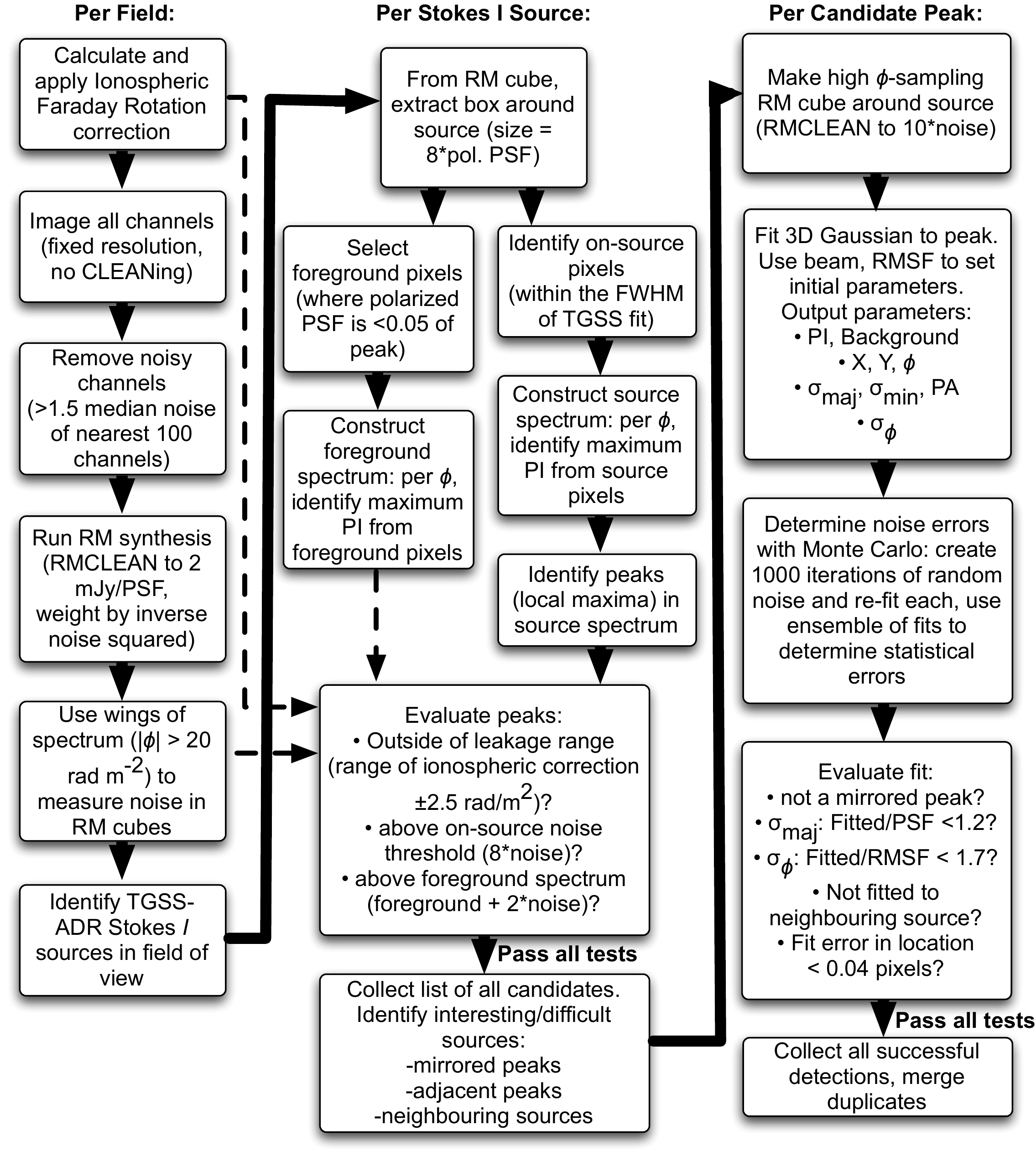} 
   \caption{A flowchart showing the key steps in the data processing. Details of each step are given in the text.}
   \label{fig:flowchart}
\end{figure*}

We used the calibrated data from the LOFAR Two-meter Sky Survey \citep[LOTSS;][]{Shimwell17}; full details on observational parameters and data calibration can be found in their paper.  The polarization calibration and imaging followed the methods used in \citet{VanEck17}, and are summarized below; a flowchart showing the overall procedure of data processing and source identification is shown in Fig.~\ref{fig:flowchart}. This paper describes the results from the 63 fields covering the Hobby-Eberly Telescope Dark Energy Experiment (HETDEX) Spring Field \citep{Hill08}, which was the region covered by the first LOTSS data release (10$^\mathrm{h}$45$^\mathrm{m}$ -- 15$^\mathrm{h}$30$^\mathrm{m}$ right ascension, 45\degr -- 57\degr\ declination). Each observation was eight hours long (giving a Stokes $I$ rms sensitivity of approximately 100 $\muup$Jy PSF\inv) with the full Dutch LOFAR array (allowing angular resolutions up to 5\arcsec\ to be achieved), covering the frequency range from 120.262 to 167.827 MHz with 488 channels (resulting in a channel width of 97.656 kHz).

\subsection{Imaging and RM synthesis}\label{sec:imaging}

We received the visibility data from LOTSS after direction-independent calibration. For ionospheric Faraday rotation correction, we used the RMextract software\footnote{https://github.com/maaijke/RMextract/} written by Maaijke Mevius, and the ionospheric total electron content (TEC) maps from the Centre for Orbital Determination in Europe (CODE)\footnote{http://aiuws.unibe.ch/ionosphere/}. The CODE TEC maps were chosen over those from the Royal Observatory of Belgium (ROB) on the basis of better results obtained in the tests made by \citet{MyThesis}. This correction introduces a systematic uncertainty in the measured Faraday depths of approximately 0.1--0.3 \radu \ \citep{Sotomayor13}.

Before imaging, we removed the baselines between the two HBA substations \citep[e.g., CS002HBA0 and CS002HBA1; see][for a description of the substation layout and naming]{vanHaarlem2013} within each station, as these are known to often have significant instrumental contamination. All the remaining core-core baselines were used for imaging. We imposed a baseline length cutoff of 800$\lambda$, to reduce the resolution so as to minimize the presence of image artifacts and to keep the resolution consistent across the full bandwidth; the resulting frequency-independent resolution was 4\farcm3, with a pixel size of 10\arcsec. The AWimager software \citep{Tasse13} was used to produce Stokes $Q$ and $U$ images for each channel. No CLEANing was performed on the channel images.

The primary purpose of the polarization imaging was the investigation of the Galactic diffuse polarized emission (which we leave to a separate paper), and so several parameters in the imaging process (particularly the baseline length cutoff and the use of short baselines) were optimized for this science goal. The resulting data products were not ideal for point source analysis (i.e., the diffuse emission dominated over the fainter point sources), but a full reprocessing using point source-optimized parameters was not possible within the scope of this work. A discussion of how the processing could be improved for polarized point source extraction (e.g., improving resolution and removing the short baselines dominated by diffuse emission) appears in Sect.~\ref{sec:discussion}.

A small fraction of the resulting images were found to have anomalously high noise levels, usually in the form of image artifacts formed by single baselines or stations with anomalously high amplitudes. To remove the affected channels, we used the non-primary beam corrected images to determine the root-mean-square noise per channel. For each channel, we compared the noise to the median noise of the 100 adjacent channels; if the noise was more than 1.5 times the median, the channel was removed and not used for the following steps. This typically removed 2--10 channels, with a few fields losing 60--80 channels. The noise level in the Stokes $Q$ and $U$ images of the remaining channels was typically 2--4 mJy PSF\inv. Three fields were found to have much higher noise levels in all channels due to the presence of polarization leakage from very bright 3C sources. These fields were dropped from the processing and did not have Faraday depth cubes made, leaving 60 fields for the remaining steps of the pipeline.

RM synthesis \citep{Burn66, Brentjens05} was performed using the pyRMsynth software\footnote{https://github.com/mrbell/pyrmsynth} developed by Mike Bell and Henrik Junklewitz. From the frequency coverage of the data, the FWHM of the rotation measure spread function (RMSF) is 1.15 \radu, and the observations have minimal sensitivity to Faraday depth structures wider than 1.0 \radu. The consequence of this is that we were not able to resolve structures in Faraday depth that are wider than the RMSF, as they would be strongly depolarized at these frequencies, and that the polarized emission we do detect must be unresolved in Faraday depth. To minimize the noise in the Faraday depth cubes we weighted the channels by the inverse of the image variance, equivalent to natural weighting in radio imaging. The nominal theoretical noise in the resulting Faraday depth cubes, assuming 480 channels with 3 mJy PSF\inv\ noise, is 0.14 mJy PSF\inv\ RMSF\inv.

For each field, Faraday depth cubes were produced that covered the Faraday depth range $|\phi|<100$ \radu, in steps of 0.5 \radu. This limited Faraday depth range was motivated primarily by data storage limitations, but was not expected to be a problem as the typical Faraday depths for sources at such high Galactic latitudes is of order a few tens of \radu; the \citet{Taylor09} RM catalog contains 910 sources in the same region as our observations, and only one source has a measured RM outside our Faraday depth range.
RM-CLEAN was applied with a threshold of 2 mJy PSF\inv\ RMSF\inv. No correction for spectral index was applied, which may introduce polarized intensity errors \citep{Brentjens05}; the degree of error has not been fully simulated for our frequency setup but is estimated to be of order 10\%.

No correction was made for the instrumental polarization leakage, as the only mitigation method developed thus far for LOFAR data \citep[SAGECal, ][]{Yatawatta2008} was deemed too computationally expensive. The polarization leakage caused part of the total intensity emission to appear as artificial polarized emission, producing a peak in the Faraday spectrum at 0 \radu\ (implying that the leakage is mostly frequency independent). However, the ionospheric correction shifted the Faraday depth of the instrumental polarization as well as the astrophysical emission. Since the ionospheric Faraday rotation ranged from 0 to 2.8 \radu \ between observations, the leakage peaks were shifted to negative values by the same amount for all the sources in the same observation while the astrophysical peaks were shifted to their true values.

To determine the noise level in the resulting cubes, we used the following method.
The expected distribution for the polarized intensity in Faraday spectrum, in the absence of signal, is a Rayleigh distribution \citep{Macquart2012, Hales2012}. For each image-plane pixel, the Faraday depths $|\phi| < 20$ \radu \ were masked out, and a Rayleigh distribution was fit to the polarized intensity distribution. The masked Faraday depth range was selected to contain most of the observed diffuse emission (as well as the polarization leakage peak) in order to remove the majority of the polarized signal. Some regions contained diffuse emission at higher Faraday depths, but this was found not to significantly affect the noise estimates. The resulting Rayleigh $\sigma$ parameter was taken as the noise in the Faraday depth spectrum for that pixel.

The noise measured with this method was observed to be position dependent in two respects. First, the noise increased smoothly with distance away from the phase center in each field, due to the station beam. Second, the noise was observed to be higher at the location of bright Stokes $I$ sources. This suggests that even though the polarization leakage is mostly confined to Faraday depths near zero, it still contributes contamination even in the wings of the Faraday spectrum. The result of this is that the on-source noise is significantly higher than the off-source noise, and that position-dependent noise estimates are necessary to properly characterize these data.

\subsection{Source candidate identification}\label{sec:candidates}
After the Faraday depth cubes were produced, the next step was to search for polarized source candidates. Source-finding directly on the polarization data was considered and rejected; a discussion on the problems of source-finding in polarization (which include non-Gaussian noise and the resulting bias in polarized intensity measurements) can be found in Farnes et al. (submitted). Instead, we chose to search for polarization only at the locations of known Stokes $I$ sources. At that time, the LOTSS catalog \citep{Shimwell17} was not available, so we used the Tata Institute of Fundamental Research (TIFR) Giant Metrewave Radio Telescope (GMRT) Sky Survey first alternative data release (TGSS-ADR1) catalog of sources \citep{Intema2017}, as this was the most sensitive catalog available at the same frequency.

At this stage, the problem of diffuse foreground contamination was considered. Diffuse polarized foreground was seen at nearly all positions in the Faraday cubes, at levels of up to 10 mJy PSF\inv\ RMSF\inv, making it necessary to develop some method of removing it or otherwise preventing it from being spuriously associated with the background sources. We first considered foreground subtraction methods, where the foreground contribution is calculated from neighbouring off-source pixels and subtracted from the on-source Faraday spectrum to give the Faraday spectrum of just the source. These methods were rejected, as we found that the foreground varies significantly on the scale of the PSF and it was not possible to accurately estimate the on-source foreground contribution. We instead chose to use foreground-thresholding: we used the neighbouring off-source pixels to measure the maximum strength of the foreground in that region (as a function of Faraday depth), and required that the on-source polarization be greater than the foreground plus twice the noise.

For each field, a list of TGSS-ADR1 sources was generated. For each source, a box centered on the source was extracted from the Faraday depth cube, with a size of 8 $\sigma_{\mathrm {maj}}$ for each axis in the image plane and covering the full Faraday depth range of the cube, where $\sigma_{\mathrm {maj}}$ defines the width of the semi-major axis of the image-plane PSF (expressed as a Gaussian $\sigma$). The box width is equivalent to 3.4 times the FWHM of the PSF. Sources too close to the edge of the cube for this box to be extracted were discarded.

The TGSS-ADR1 fitted source size was overlaid on this box, and pixels within the source FWHM were classified as `on-source' pixels for the next step.\footnote{After processing we realized this was not ideal, as there may be extended sources where only parts of the source, including parts outside the fitted FWHM, are polarized. In principle, such sources might not be detected if the polarization was far enough from the selected `on-source' pixels and quite weak (enough that our PSF would not move enough polarized signal into the `on-source pixels'). However, given our very coarse angular resolution and the small number of sources extended enough to be affected (only 27 out of 13218 TGSS sources in this region have fitted sizes larger then 2\arcmin), we estimated that it was unlikely that we missed a source due to this effect and decided it was not necessary to develop a new method and reprocess all the observations. Future polarization pipelines, particularly those at higher angular resolution, should more carefully consider solutions to this problem.}
A foreground mask was constructed by overlaying the polarized PSF at the location of each TGSS-ADR1 source in or nearby the box; this was to prevent polarized components from the target source or neighbouring sources from being considered as foreground. All pixels below 5\% of the PSF peak, approximately 60\% of the box if no neighbouring sources were present, were classified as foreground pixels for the next step.

\begin{figure}[!h]
   \centering
   \includegraphics[width=\linewidth]{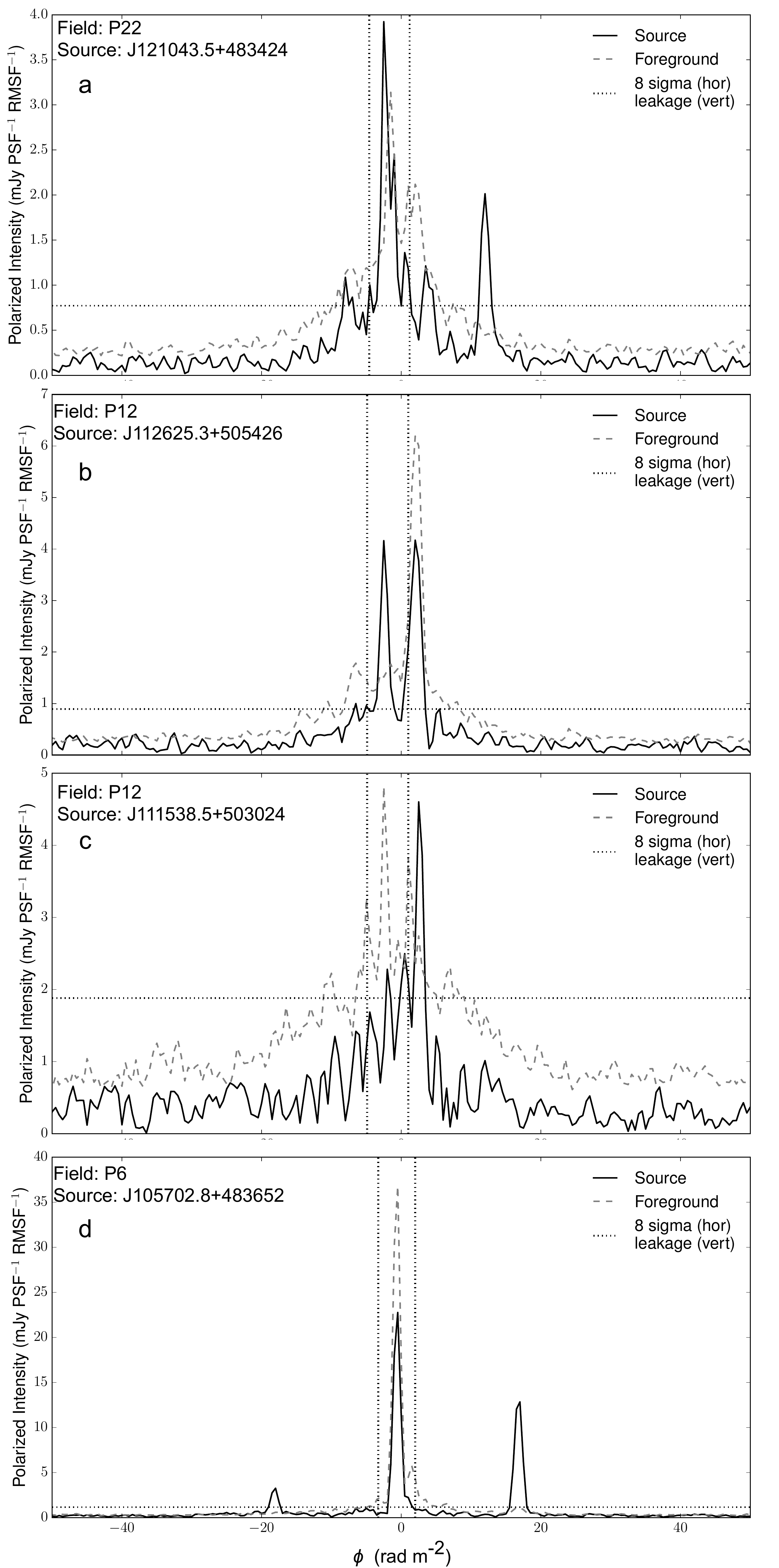} 
   \caption{Four examples of Faraday depth spectra of different sources. For each, the source's Faraday spectrum and foreground Faraday spectrum are shown in black and grey respectively, the 8-sigma (noise) level is marked in a horizontal dotted line, and the excluded Faraday depth range around the instrumental leakage is bounded by vertical dotted lines. The four spectra were selected to show examples of different phenomena. {\it a)} a typical detection that passed all of the tests in Sect.~\ref{sec:evaluation}, with a clear bright source at +12 \radu; {\it b)} A non-detection with bright foreground (both on-source and off-source) at +3 \radu, showing the need to consider the foreground emission when identifying sources; {\it c)} a polarized candidate (which failed the tests) at +3 \radu, which can be seen in the full 3D cube to be a local enhancement in the foreground emission; {\it d)} a polarized source at +17 \radu \ that passed all tests, with an artificial `mirrored' peak at -18 \radu \ (an artifact discussed in Sect.~\ref{sec:evaluation}).}
   \label{fig:spectra}
\end{figure}

The source Faraday depth spectrum was constructed by taking the maximum polarized intensity of the on-source pixels at each Faraday depth. The foreground spectrum was constructed in the same way for the foreground pixels around the source. The source spectrum was then searched for peaks, simply by identifying all the local maxima in the spectrum. Figure~\ref{fig:spectra} shows four examples of source and foreground Faraday depth spectra constructed using this method.

From the noise map calculated in the previous section, the highest noise of the on-source pixels was selected as the noise at the source location. For each peak, a series of tests was applied to select detection candidates. The first test was that a peak must have polarized intensity greater than 8 times the noise. This removed low-intensity noise-like peaks. The second was that the peak must have a polarized intensity greater than the foreground spectrum plus two times the noise. The third test was that the peak had to be clearly separated from the instrumental leakage peak. Peaks that were within 2.5 \radu \ of the range of values for the ionospheric correction for that observation were rejected. The free parameters in these tests were tuned manually to minimize the number of false detections while not removing any peaks that looked promising to a by-eye inspection. This process resulted in 795 candidate peaks; this number contains both sources appearing in multiple fields (due to overlap in the image-plane) and sources with multiple peaks in Faraday depth.

The full-field Faraday depth cubes were too coarsely sampled in Faraday depth for high-accuracy measurements, so for each source with one or more candidate peaks a new postage-stamp Faraday depth cube with high oversampling in Faraday depth was produced. These cubes had image-plane dimensions equal to the size of the box used in the previous step, and spanned the Faraday depth range $\pm100$ \radu \ in steps of 0.1 \radu. RM-CLEAN was performed to a threshold of 10 times the source noise. From these Faraday cubes, the Faraday depth range $\pm 2$ \radu\ around each candidate peak was extracted and used to fit the peak.

The fitting function used was a 3D Gaussian, formed by multiplying a general 2D image-plane Gaussian with a 1D Faraday depth Gaussian. This was chosen to match the expected form of an unresolved source: the PSF and RMSF were well approximated by a Gaussian, and a Gaussian restoring function was used in the RM-CLEAN algorithm. A non-zero background level was also included as a free parameter, added in quadrature with the Gaussian. Quadrature addition was chosen as it was expected that the noise and foreground would add to the source signal as complex components while we were fitting to polarized intensity only. This resulted in a nine-parameter model for each peak: a peak polarized intensity ($P$), a background polarized intensity ($C$), the image-plane centroids ($X$ and $Y$, in pixel coordinates), the image-plane semi-major axis ($\sigma_\mathrm{maj}$), semi-minor axis ($\sigma_\mathrm{min}$), and position angle (PA), the Faraday depth centroid ($\phi$), and the Faraday depth width ($\sigma_\phi$).\footnote{The three width parameters were kept as Gaussian $\sigma$, rather than expressed as FWHM. This convention will be followed throughout the paper unless otherwise specified.}

We performed the fitting using the `curve\_fit' task from the Python SciPy optimize module, which uses a Levenberg-Marquardt least-squares algorithm.
The initial guess parameters, in the same order as above, were the peak polarized intensity from the full-field cube, the source noise, the TGSS-ADR1 source location, the image-plane PSF size and orientation, the peak Faraday depth from the full-field cube, and the RMSF width. For each peak, the best-fit parameters and the fit errors reported by curve\_fit were recorded. Peaks where the fitting algorithm failed to converge were labeled as false detections and discarded; this accounted for 47 of the 795 candidates. These candidates were investigated and found to be mostly comprised of weak Faraday-depth sidelobes of the instrumental leakage with very high skew, with the rest being cases of extremely spatially-complex diffuse emission. We inspected sources with failed fits visually, and concluded that none of these candidates can be real sources.

\subsection{Error analysis}\label{sec:errors}
The fit errors produced by curve\_fit were not reliable, as they were unrealistically small in nearly all cases (for example, typical errors in centroid position of a few hundredths of a pixel). These errors are calculated assuming that all data points are uncorrelated; however, significant correlation structure is present in our Faraday depth cubes due to the limited resolution, as defined by the PSF and RMSF.
The result is that the number of effective degrees of freedom is much smaller than what is assumed, so the true errors are much larger than the those derived by curve\_fit.

To derive more reasonable errors, we attempted a different approach, using Monte-Carlo techniques. Bootstrapping methods were considered and rejected, as a naive approach to bootstrapping (such as simply randomizing the fit residuals) would not reproduce the correlation structure of the noise. Instead, we chose to create randomized realizations of noise with correlation structure as similar to the real noise as possible, and simulate the fitting procedure on these noise realizations.

To produce simulated noise with the correct characteristics, we considered the source of the noise in the data. During the imaging process, the input visibilities contain noise, so each grid-point in the ($u,v$) plane that contains visibilities will also have an associated noise, which should be independent of other points (neglecting the effects of a convolution kernel in the image gridding process, which we do not expect to be significant). Unfortunately, the imager software does not provide access to these noise values. Without this, the only information we had on the correlation structure in the data is the PSF. We performed an FFT of the 3D PSF+RMSF to recover the distribution of Fourier components in the PSF. Each component was then multiplied by a random complex number with both real and imaginary components drawn from a Gaussian distribution with zero mean and a standard deviation of 1. The resulting randomized Fourier components were transformed back into a Faraday depth cube and the imaginary component discarded, producing a randomized noise realization with correlation structure that appeared very similar to the real data. The amplitude of the noise realization was then scaled so that the standard deviation matched the source noise.

For each source, 1000 such random noise realizations were created. Each realization was added to the best-fit model, a new fit was performed, and the resulting fit parameters were recorded. For each fit parameter, the standard deviation of the 1000 simulated fits was used as the estimated error caused by noise. For all parameters, the noise error was typically 30--100 times greater than the fit error, so the noise errors are taken to represent the true uncertainties on the fit parameters (see Sect.~\ref{sec:verification} for tests of this assumption). For some parameters, particularly the fitted size parameters, the simulated distributions were observed be significantly non-Gaussian, and consequently that the standard deviation does not completely describe the underlying statistics; however, the key parameters for the catalog (position, polarized intensity, and Faraday depth) all had distributions that appeared very Gaussian-like to visual inspection.

There are at least three issues with this method that we were not able to address. First, the use of the PSF and RMSF to determine the Fourier components of the noise is not correct. The Fourier transform of the PSF/RMSF should be equal to the weights used in the imaging/RM synthesis processes, rather than the noise. The weights will not be proportional to the noise, and for some weighting schemes (e.g. natural weighting) will actually be anti-correlated with the noise value for each Fourier component. For this reason, the PSF/RMSF will show which Fourier components are present in the data, but will not give the correct relative amplitudes. 

Second, since this method combines the noise realization with the best-fit model to produce a new simulated dataset, it does not take into account any uncertainties that result from unfitted structure in the real data (i.e. in the residuals). Any differences between the data and the source model, such as irregularly shaped resolved sources, which could affect the estimation of fit parameters are not accounted for in this process.

Third, the resulting noise, and its combination with the best-fit model, do not follow the proper statistical distributions for polarized intensity. The simulated noise has a Gaussian distribution, while polarized intensity should follow a Ricean distribution. For high signal-to-noise cases the current method should be very close to accurate (as the Ricean distribution becomes more Gaussian-like at high signal-to-noise), but for lower signal cases and for the off-source pixels the difference will be greater. A more careful treatment using the full complex polarization for both the source model and the noise realizations would resolve this problem.

Despite these unresolved problems, we consider the resulting uncertainties to be the most accurate estimation available for the true uncertainties in our measured fit parameters; in Sect.~\ref{sec:verification} we evaluate the quality of the errors produced from this method.

\subsection{Candidate evaluation}\label{sec:evaluation}
A large number of the candidate peaks were clearly not real polarized sources (e.g., were indistinguishable from adjacent foreground emission, were clearly separated in position from the Stokes $I$ source, or appeared to be sidelobes of the instrumental leakage), so additional selection tests were necessary to separate the reliable detections from the probable false detections.

To identify polarized sources in an automated way, we compiled a series of criteria that we describe in the next few paragraphs. We tested the effectiveness of these criteria on a catalogue of candidates that we classified by eye based on whether they are isolated in the 3D Faraday depth cube and whether they appear to be genuine, clearly fake, or ambiguous. Of the candidates that were isolated in the Faraday depth cube, 129 were classified as real, 433 as false, and 62 as ambiguous. Of the candidates with neighbouring candidates, 85 were classified as real, 65 as false detections, and 21 as ambiguous.

The first test was to remove candidates that were `mirrored peaks', a type of instrumental artifact sometimes seen with pyRMsynth where a bright real peak produced a weaker peak on the opposite side of the leakage peak. 
Some Faraday spectra that contained a bright real peak and a strong leakage peak would have a third, weaker peak on the opposite side of the leakage peak from the real peak and with the same separation in Faraday depth, giving the appearance of being `reflected' across the leakage peak. The bottom panel of Fig.~\ref{fig:spectra} shows an example of such a Faraday spectrum. We checked each candidate for such mirrored peaks; in those cases we discarded the peak with the weaker polarized flux density. 

\begin{figure}[!h]
   \centering
   \includegraphics[width=\linewidth]{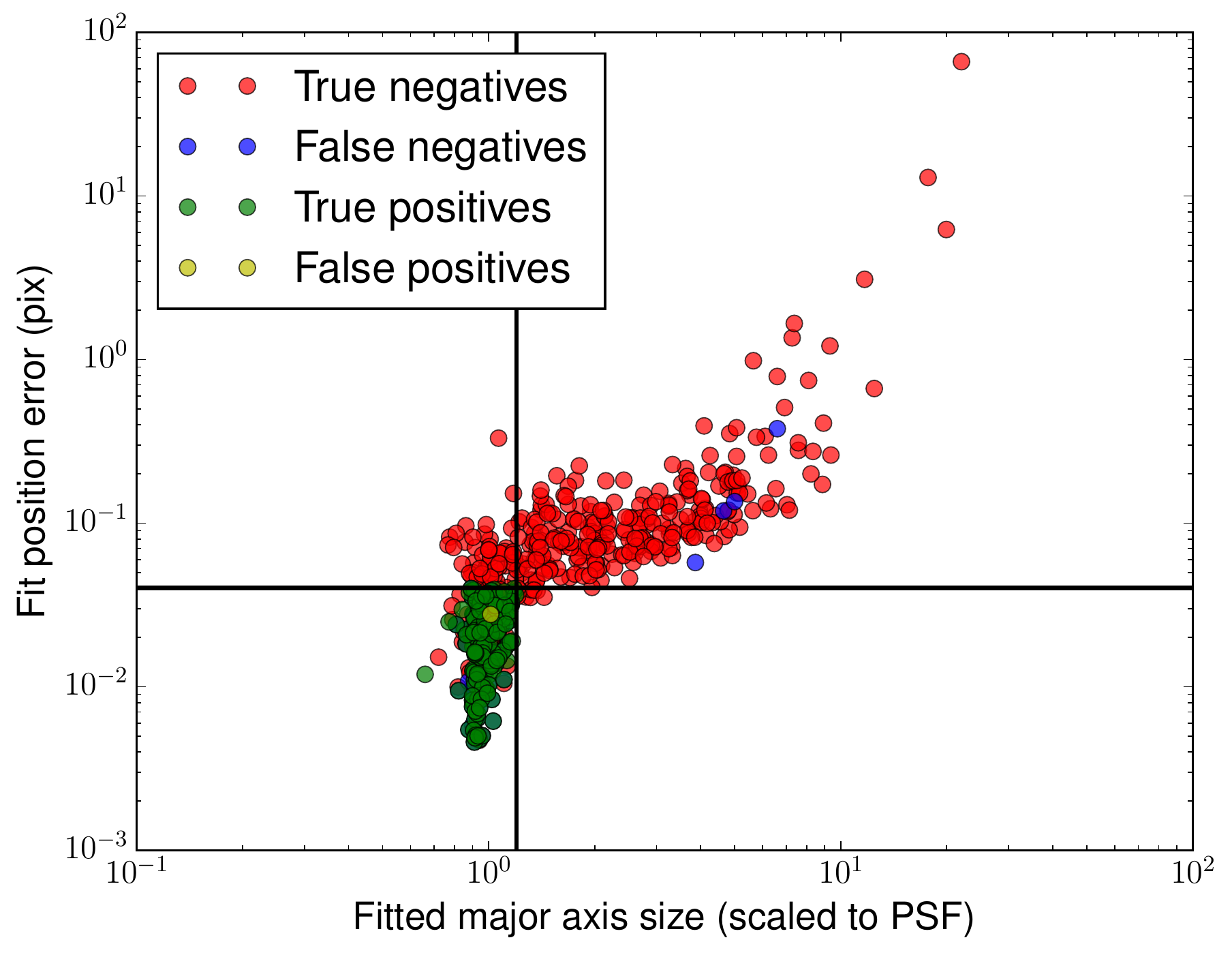} 
   \caption{Two of the tests to determine which candidates were real: the size of the fitted major axis relative to the PSF (horizontal axis) and the error in position from the fit (vertical axis). The black lines show the test thresholds: 1.2 times the fitted major axis size and 0.04 pixels position error. Each point is a candidate peak, colored by how it was evaluated by the tests (positives passed all tests, negatives failed one or more tests, including the tests not shown) and the manual inspection (true positives/negatives where the tests match the manual inspection, false where the tests do not match the manual inspection).}
   \label{fig:tests}
\end{figure}

The next two tests were to discriminate between unresolved point-sources and sources that appeared extended in either the image plane or in Faraday depth. Due to the poor image plane resolution of the polarization data and the strong presence of diffuse foreground, we made the assumption that any fit that deviated significantly from the PSF was much more likely to be a diffuse polarized feature than an extended polarized background source. The properties of the RMSF meant that we were not sensitive to resolved structures in Faraday depth, so fits significantly broader than the RMSF were interpreted as artifacts, as they were often found to be sidelobes of the instrumental leakage. To determine which parameters and thresholds were effective in discriminating between real and false sources, we compared the distributions of the sources identified by-eye as isolated real sources to those identified as isolated false detections. We found that the fitted major axis ($\sigma_\mathrm{maj}$) and Faraday depth width ($\sigma_\phi$) were the most effective discriminants. We found that the fitted minor axis and position angle were not useful to distinguish between real and false detections. Sources with $\sigma_\mathrm{maj}$ greater than 1.2 times that of the PSF (vertical line in Fig.~\ref{fig:tests}) or with $\sigma_\phi$ greater than 1.7 times that of the RMSF were rejected.

Next, we removed duplicate candidates within the same observation, which were caused when multiple TGSS-ADR1 sources were close to each other (typically within about 2\arcmin, or about half the PSF FWHM), but only one was polarized; the neighbouring sources would also be classified as candidate polarized sources and go through the fitting procedure. Since these detections represented duplicates produced from the same data, differing only by which TGSS-ADR1 source they were associated with, these duplicates were removed. The exact selection criterion was to remove the detections where the fitted polarization position was closer to a different TGSS-ADR1 source; this left only the detections where the TGSS-ADR1 source position was closest to the polarization position.


We found that several of the fit errors were very powerful discriminants between real and false candidates. As described in the previous section, these errors were unphysical, but they appeared to still be sensitive to the quality of the fit. Our interpretation is that large fit errors are likely caused when significant non-Gaussianities are present, and this is often a sign that the polarized emission being fitted is diffuse Galactic emission and not from the point source. After looking at the different fit errors, the most effective test appeared to be the position error (the X and Y centroid errors added in quadrature). Candidates with position errors larger than 0.04 pixels were rejected. Fig.~\ref{fig:tests} shows how effective this test was in discriminating between candidates that did not pass the manual inspection; the fitted position error and the major axis were the most effective tests, together removing 80\% of the false candidates.

After these tests, we found that 114 of the 129 isolated real candidates and 1 of the 433 isolated false candidates passed all tests. The tests described above were applied to all candidate peaks, resulting in 177 passing candidates (hereafter called detections). Due to the overlap between adjacent fields many sources were detected multiple times, so the number of unique TGSS-ADR1 sources with one or more passing detections is 92. None of the sources were found to have more than one Faraday depth component that passed all tests.

\subsection{Catalog verification}\label{sec:verification}
Due to the partial overlap between observations, most of the 92 sources in our catalog were identified as candidates multiple times in independent observations, which offered an opportunity to verify the reliability of the measurements and our pipeline. From the catalog, 33 sources were candidates in two fields, 23 in three fields, and 13 in four fields. 

\begin{figure}[!h]
   \centering
   \includegraphics[width=\linewidth]{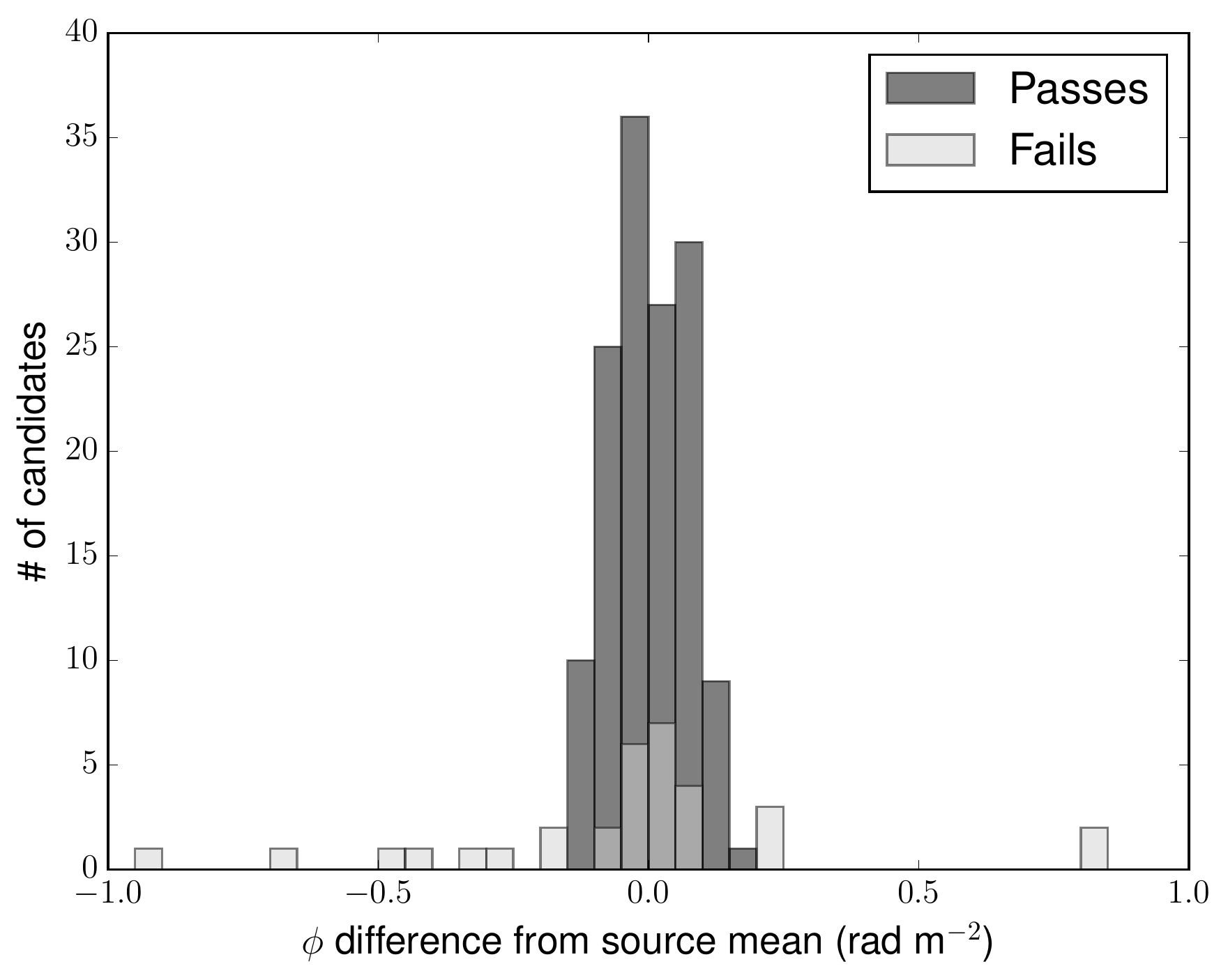} 
   \caption{Histograms of the mean-subtracted residual Faraday depth for sources with multiple candidates from different observations, where the mean is calculated using only those observations that passed the quantitative tests (the detections). The candidates are separated by whether they passed the quantitative tests or not (each source must have at least one detection to be included).}
   \label{fig:phi-differences}
\end{figure}

To assess the consistency of the Faraday depth measurements, we looked at the variations in the fitted Faraday depth, using only the sources that passed the quantitative tests at least once (but including the non-passing candidates of those same sources). For each catalog source that was observed multiple times we calculated the mean Faraday depth of the observations, and we analyzed the residual Faraday depths after subtracting this mean from the observed values. Figure \ref{fig:phi-differences} shows the distributions of these residuals.
This clearly shows that the candidates that did not pass the tests often show much larger Faraday depth variations than the detections.

Using the differences shown in Fig.~\ref{fig:phi-differences}, we also estimated the variations in measured Faraday depth introduced by the ionospheric Faraday rotation correction. The differences between the measured Faraday depth for different observations of the same source are due to measurement uncertainty and inter-night variations in the ionospheric correction, which together contribute to the width of the distribution in Fig.~\ref{fig:phi-differences}. However, there were a few sources where the ionospheric correction would not contribute to the difference: in some cases adjacent fields were observed simultaneously, meaning that they were observed with identical ionospheric conditions and ionospheric Faraday rotation corrections. These sources were removed from the sample used to calculate the ionospheric correction variability below.

The standard deviation of the Faraday depth residuals is 0.071 \radu, considering only the detections that passed the quantitative tests (Fig.~\ref{fig:phi-differences}). The root-mean-square noise error in $\phi$, as estimated from Sect.~\ref{sec:errors}, for the same detections is 0.047 \radu. The two sources of variability (noise and ionospheric correction) are statistically independent, so subtracting the noise contribution from the observed standard deviation in quadrature yields an estimated correction variability of 0.053 \radu. This is smaller than the 0.1--0.3 \radu\ uncertainty estimated by \citet{Sotomayor13}. We interpret this difference to mean either the ionospheric correction uncertainty is smaller than expected, or a significant portion of the uncertainty is systematic and affects all observations equally (such a systematic offset would be removed by the differencing used to produce Fig.~4).

\begin{figure}[!h]
   \centering
   \includegraphics[width=\linewidth]{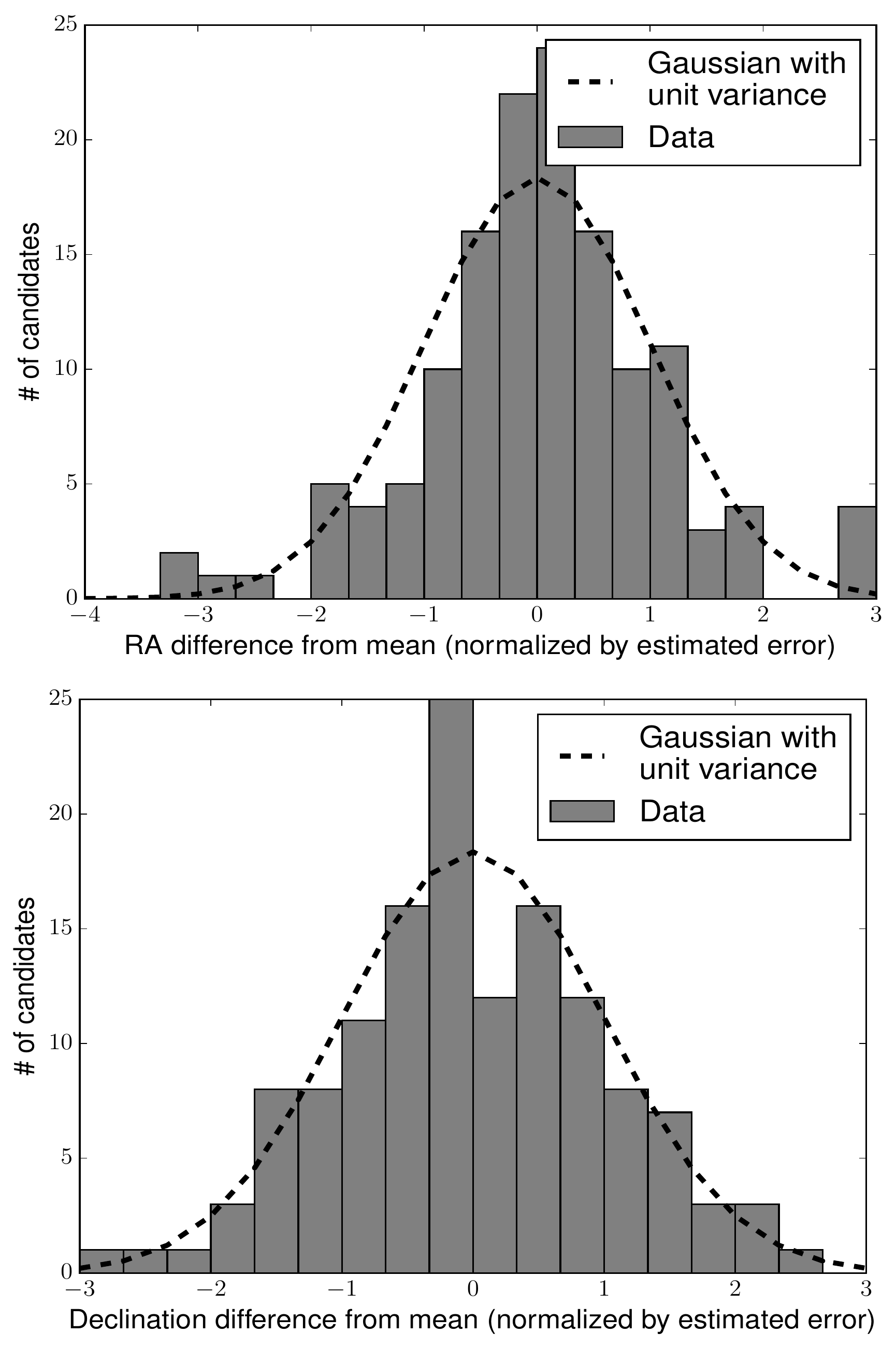} 
   \caption{The distributions of offsets in position (right ascension ({\it top}) and declination ({\it bottom})) from the mean for multiply-detected sources. For each source with 2 or more detections, the mean-subtracted residuals of position for each detection were calculated. The residuals were normalized by the errors calculated from the Monte Carlo noise simulations, and the resulting distributions are shown. The dashed lines are the expected distribution if the errors are correctly calculated: normal distributions with a variance of 1.0 and normalized to the total number of detections.}
   \label{fig:position_errors}
\end{figure}

To test whether the measurement uncertainties we estimated in Sect.~\ref{sec:errors} were realistic, we performed a similar analysis as above, but replacing Faraday depth with right ascension (RA) and declination. These were chosen as they were expected to not suffer from any polarization-specific complications, and during the noise simulations they were observed to have a Gaussian-like distribution.
For sources that were observed multiple times we calculated the mean RA and declination and the uncertainties in these parameters, including only those observations that passed all the criteria listed in section 2.4. These mean values divided by the estimated uncertainty in the mean are shown in Fig.~\ref{fig:position_errors}, and are expected to follow a standard normal distribution.
Qualitatively, both distributions are well represented by a Gaussian with unit variance. We calculated the standard deviation for each distribution as 1.08 for RA and 1.09 for declination, which suggests that the noise errors we calculated may be underestimated by a few percent. Since this difference is quite small, we chose to leave the errors as calculated, with no rescaling.

The final step to produce the catalog was to combine the multiple detections into single catalog entries. For each source with multiple detections, the positions, polarized intensities, and Faraday depths were averaged using only those detections that passed all the tests described in Sect.~\ref{sec:evaluation}.

\section{Polarized source catalog}\label{sec:catalog}
The processing and selection steps described in the previous section resulted in a catalog of 92 polarized sources. One source matched to the pulsar B1112+50 in the ATNF pulsar catalog \citep{ATNF}\footnote{http://www.atnf.csiro.au/people/pulsar/psrcat/}. This pulsar has a previously measured RM of $-0.1\pm0.8$ \radu\ \citep{Force2015}, while we measured $+2.69\pm0.01$; \citet{Force2015} do not report applying any ionospheric Faraday rotation corrections, so this may be the cause of the difference.

\begin{figure*}[!h]
   \centering
   \includegraphics[width=\linewidth]{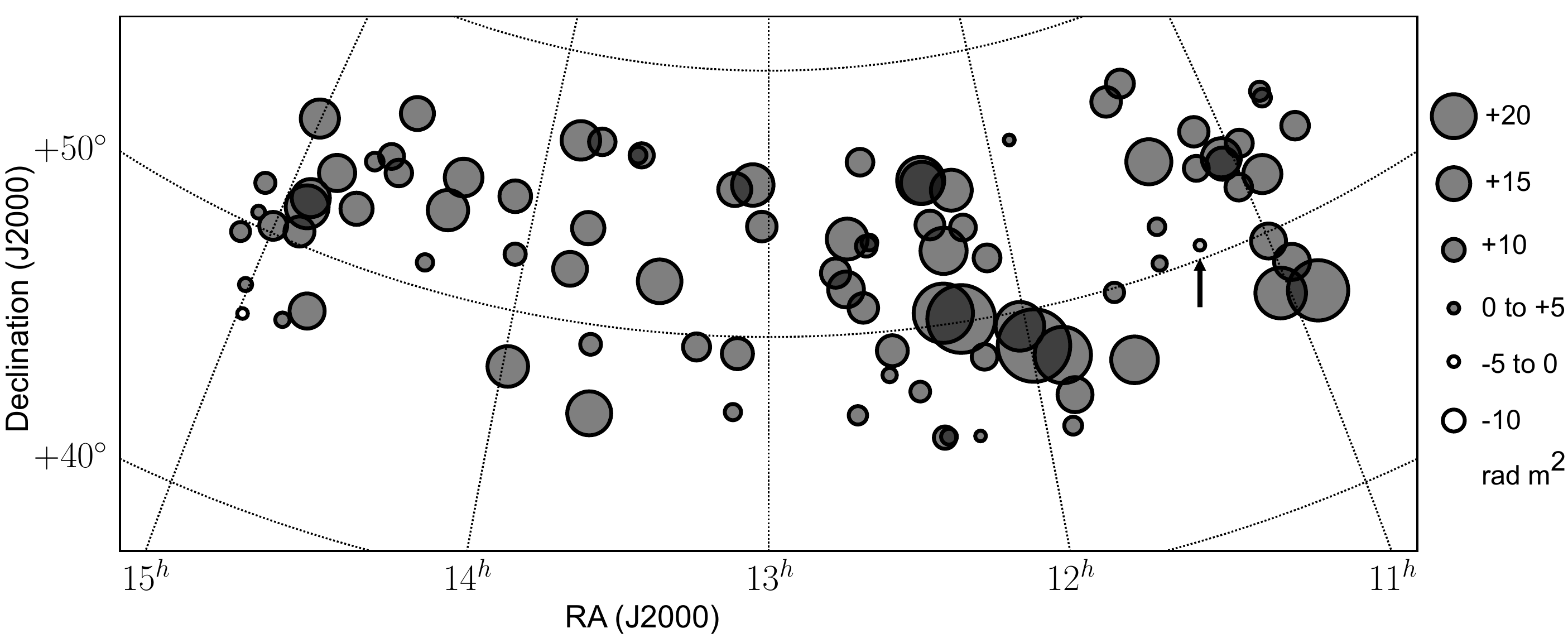} 
   \caption{The positions and Faraday depths of all polarized sources in the catalog. The size of each symbol is proportional to the magnitude of Faraday depth (sources with $|\phi| < 5$\ \radu\ are set equal in size to 5 \radu; the largest circle is 32 \radu) while open and filled circles represent negative and positive Faraday depths respectively. The single pulsar (B1112+50) is marked with an arrow.}
   \label{fig:RM_positions}
\end{figure*}

The catalog is presented in Table \ref{tab:catalog}, with the pulsar moved to the end. The RMs from matching sources in the 1.4 GHz \citet{Taylor09} catalog are included for reference. Figure \ref{fig:RM_positions} shows the positions and Faraday depths of these sources; the total area covered is approximately 570 square degrees. The position, polarized flux density, and Faraday depth of each source were taken directly from the 3D fit parameters described previously, except for sources with multiple detections where the parameters from these detections were averaged as described above. The errors in each of these parameters were taken to be the noise errors calculated previously. No correction for polarization bias \citep{Simmons85} was applied, as it was not clear how our fitting procedure (particularly the inclusion of the offset term) was affected by polarization bias. The Faraday depth measurements will have an additional error contribution, in addition to the error values reported in the table, introduced by the ionospheric Faraday rotation correction, which was estimated in Sect.~\ref{sec:verification} as about 0.05 \radu.

\section{Analysis}\label{sec:analysis}
Below we present various types of analysis that can be made using the values from our catalog.

\subsection{Comparison with NVSS rotation measures}
\begin{figure}[!h]
   \centering
   \includegraphics[width=\linewidth]{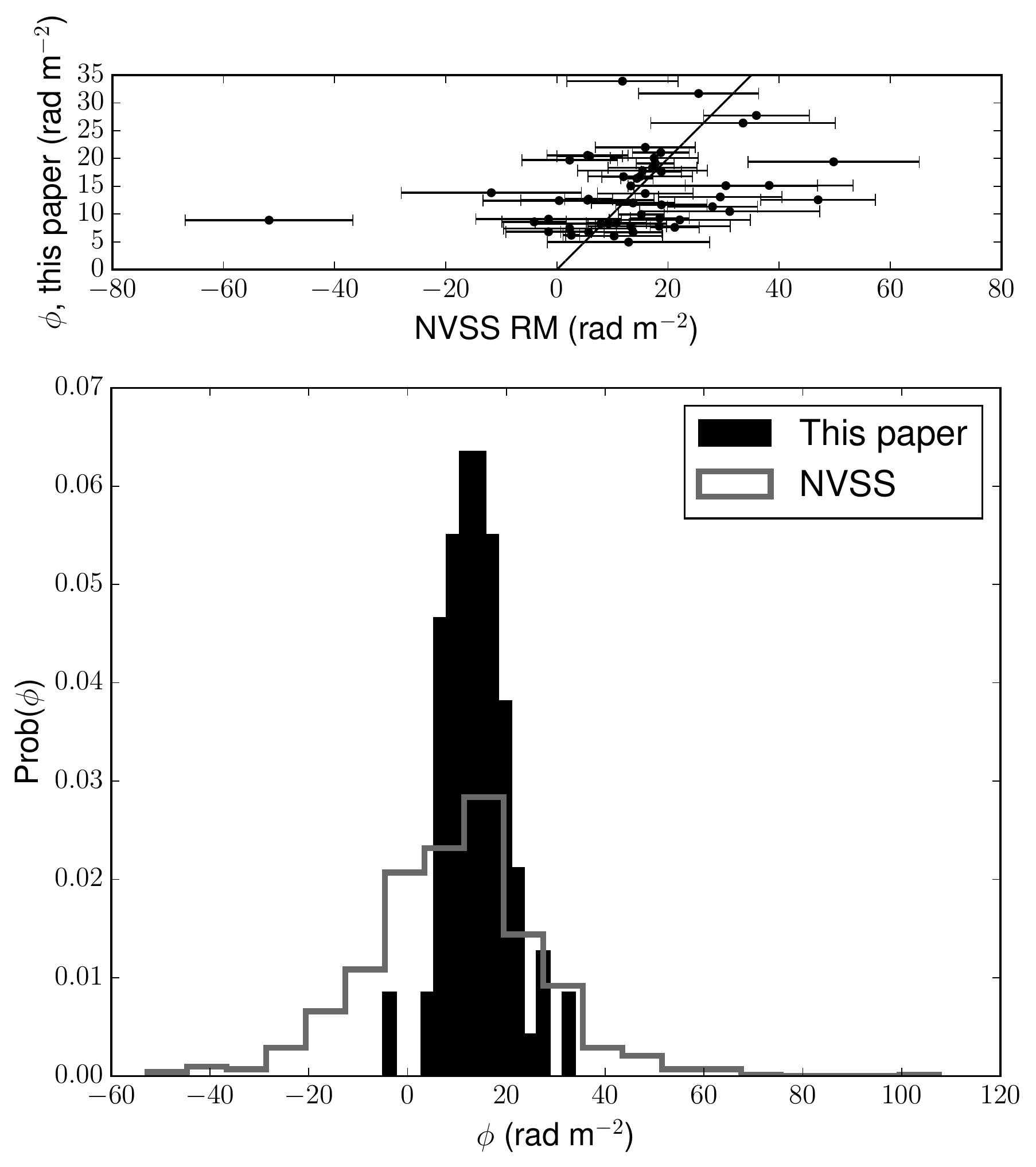} 
   \caption{{\it Top:} A comparison of the measured Faraday depths for the 51 sources present in both our catalog and the \citet{Taylor09} catalog. The diagonal line marks 1:1 correspondence. The errors in our measured Faraday depth are almost always much smaller than the symbol size.  
   {\it Bottom:} The distribution of Faraday depths for our catalog and the \citet{Taylor09} catalog. The absence of sources near 0 \radu \ in our catalog is due to our procedure of ignoring the Faraday depths very close to the instrumental leakage.}
   \label{fig:RM_distribution}
\end{figure}

To compare our low-frequency Faraday depth measurements to higher frequency measurements, we cross-matched our sources against those in the catalog of \citet{Taylor09}, which used 1.4 GHz observations. Their catalog had 910 sources in the same area as ours. Using a cross-matching limit of 1 arcminute, we identified 51 sources that appeared in both catalogs. The top panel of Fig.~\ref{fig:RM_distribution} compares the measured Faraday depths between the catalogs. The two catalogs are in approximate agreement, with a large scatter. The bottom panel of Fig.~\ref{fig:RM_distribution} shows the distribution of Faraday depths for all sources in the HETDEX Spring Field region from each catalog. Our catalog has a much narrower distribution, likely due to the smaller errors on the Faraday depth measurements, and a notable absence of sources near 0 \radu, which is due to candidates near to the instrumental leakage (typically between 0 and -2 \radu, due to the ionospheric correction) being deliberately excluded in our analysis. As a result, our catalog is almost certainly incomplete, and biased against Faraday depths near zero \radu.

We performed a chi-squared test of the difference in Faraday depth between the two catalogs, and found a reduced chi-square statistic of 3.9 (indicating a root-mean square residual of about 2 $\sigma_\phi$), suggesting that the scatter is significantly larger than we would expect from the errors. While there are some suggestions that the errors in the \citet{Taylor09} catalog are underestimated \citep{Stil11}, that is not expected to be significant enough to cause this. One plausible explanation is that many of these sources possess some Faraday-thick (by LOFAR standards) polarized emission, which would be strongly depolarized at LOFAR frequencies but could still contribute at 1.4 GHz. Many background polarized sources show this sort of broad Faraday structure, but it requires very broad bandwidth observations, including much higher frequencies, to measure such structure \citep{Anderson16}.

\subsection{Polarized source surface density}
While our catalog is incomplete, due to the effects of the instrumental leakage, beam depolarization, and strong foregrounds, we can still estimate a lower limit for the number density of polarized sources at 150 MHz. The total area covered by our observations, accounting for overlap between fields, is approximately 570 square degrees, with a typical 5-sigma sensitivity of 1 mJy PSF\inv\ RMSF\inv. Therefore, without making any corrections for incompleteness, the 92 polarized sources (including the pulsar) in our catalog give a polarized source density of 0.16$\pm$0.02 sources per square degree, or 1 source per 6.2 square degrees. 

For comparison, \citet{Mulcahy14} found 6 polarized sources in a single LOFAR observation, at 20\arcsec\ resolution, 100 $\muup$Jy PSF\inv\ RMSF\inv\ sensitivity, and an area of 17.3 square degrees, giving a polarized source density of 0.35$\pm$0.14 sources per square degree, or one source per 2.9 square degrees.\footnote{\citet{Mulcahy14} give their source density as one per 1.7 square degrees, which is a higher density than the direct calculation above. They may have made a correction for the reduced sensitivity at the edges of the primary beam, but this is not described in their paper.} \citet{VanEck17} reported 3 polarized sources in their field centred on IC~342, and \citet{Jelic15} reported 16 in their 3C~196 field, with observations of similar depth and area to \citet{Mulcahy14} but lower resolution (4.5\arcmin\ and 3.9\arcmin, respectively). Neld et al. (submitted) found 6 sources (at a confidence level of 95\%) in a 19.6 square degree region (one source per 3.2 square degrees)

\citet{Bernardi13} searched for polarized sources with the MWA at 189 MHz, and found 1 source in 2400 square degrees, but with a much higher polarized flux density threshold of 200 mJy PSF\inv\ (for comparison, we found no sources above this threshold). \citet{Bernardi13} also calculated, from previously published 350 MHz polarized source detections, that the typical source density at 350 MHz is roughly 1 source per 4 square degrees at a sensitivity of 3--12 mJy PSF\inv. Given the (unquantified) incompleteness of our sample, our source density may be in general agreement with this prediction.

\subsection{Average magnetic field}
Subject to certain caveats, it is possible to estimate the average magnetic field parallel to the line of sight by using the relationship between the Faraday depth ($\phi = \int n_\mathrm{e} B_\parallel dl$) and the dispersion measure (DM = $\int n_\mathrm{e} dl$). Specifically, under the assumption that the magnetic field and the free electron density are statistically uncorrelated, the electron density-weighted average parallel magnetic field is defined as $\langle B_\parallel \rangle = 1.232\, \phi / \mathrm{DM}$ \citep{Beck2003}.

For the single pulsar, B1112+50, the reported DM is 9.18634$\pm$0.00026 pc cm$^{-3}$ \citep{Bilous16}, and the measured Faraday depth\footnote{We have linearly added an additional error contribution of 0.05 \radu\ to the measurement error of 0.01 \radu, to account for uncertainty in the ionospheric correction.} is +2.69$\pm$0.06 \radu, giving an estimated $\langle B_\parallel \rangle$ of 0.361$\pm$0.008 $\muup$G. From the YMW16 electron density model \citep{YMW16}, the estimated distance of this pulsar is 0.97 kpc.

For the remaining sources, we estimate the Galactic DM contribution using the YMW16 electron density model. This model predicts DMs between 20 and 24 pc cm$^{-3}$ for lines of sight in this region, integrating out to 30 kpc. We assume our sources are extragalactic, aside from the known pulsar, and that the DM contributions are negligible beyond 30 kpc. For comparison, the NE2001 model \citep{NE2001} predicts DMs between 28--31 pc cm$^{-3}$. We chose to take the typical dispersion measure to the edge of the Milky Way in this region as 25$\pm$5 pc cm$^{-3}$. The Faraday depth distribution of our polarized sources has a mean of 12 \radu\ and a standard deviation of 7 \radu. We assume that the extragalactic contribution to the Faraday depths has a statistical mean of zero and thus can be ignored. Combining these values, we estimate the average parallel magnetic field strength as 0.6$\pm$0.3 $\muup$G.

However, the assumption of statistical independence between the magnetic field and the electron density is probably not accurate, especially for a high Galactic latitude field like this. Both the free electron density and the magnetic field strength decrease with distance from the Galactic plane, which will result in a correlation even in the absence of any physical effects that might cause the two to be related. As a result, this magnetic field strength is likely most representative of regions of highest electron density, near the Galactic plane just beyond the Local Bubble.

\subsection{Fractional polarization distribution}
\begin{figure}[!h]
   \centering
   \includegraphics[width=\linewidth]{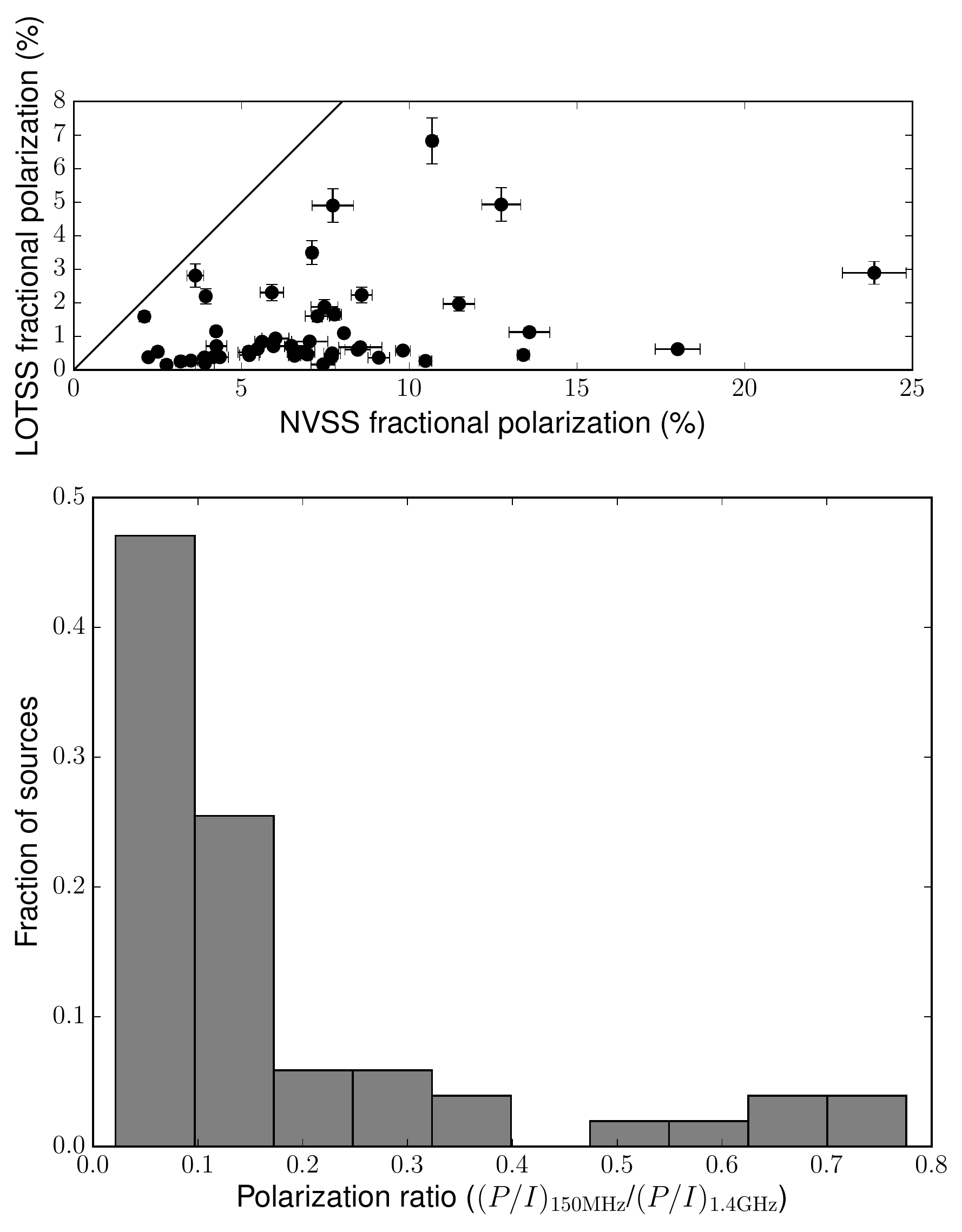} 
   \caption{{\it Top:} A comparison of the fractional polarization (ratio of polarized flux density to total intensity) at 150 MHz (our catalog, vertical axis) and 1.4 GHz (\citet{Taylor09} catalog, horizontal axis), for the 51 sources present in both catalogs. The black line marks 1:1 equivalence.
   {\it Bottom:} Histogram of the depolarization ratio, which is defined as the ratio of 150 MHz fractional polarization to 1.4 GHz fractional polarization, for the same sources.}
   \label{fig:fracpol}
\end{figure}

The frequency dependence of the fractional polarization can be used to measure depolarization and in turn learn about Faraday depth structure inside polarized sources \citep[e.g.,][]{Farnes14,Lamee2016}.
To investigate the frequency-dependent depolarization processes in our sources, we compared the polarization fractions of our data with those of the matching sources in the \citet{Taylor09} catalog. To calculate the fractional polarization, we took the ratio of the measured polarized flux density and the integrated flux from the TGSS-ADR1 catalog. The integrated flux was chosen over the peak flux as many of the sources were resolved in the TGSS-ADR1, but none were resolved in our polarization data (due to the much coarser resolution). The top panel of Fig.~\ref{fig:fracpol} shows the resulting comparison. All of the sources seen in the LOTSS data have lower fractional polarization than at 1.4 GHz, and most are less than 1\% polarized at 150 MHz.

Since our sources are unresolved in both catalogs, we can consider both measurements as probing the (solid angle-) integrated Faraday depth spectrum of these sources. By comparing the low-frequency measurements, which are sensitive to only narrow Faraday depth components, to higher frequency data, which include both narrow and broader Faraday depth features, we can gain some information on the distribution of polarized flux in the integrated Faraday depth spectrum.

In our observations, Faraday depth structures thicker than about 1 \radu\ are strongly depolarized,\footnote{Note that the depolarization depends not just on the width of the Faraday depth distribution, but also on the shape. Where relevant, we are assuming rectangular shaped features in Faraday depth (Burn slabs \citep{Burn66} and related distributions \citep{Schnitzeler2015b}); the Faraday depth widths quoted may vary by a factor of a few if other distributions or definitions of width (e.g., Gaussian $\sigma$) are assumed.}
leaving us sensitive to only the narrow ($\lesssim$1 \radu) components; the sensitivity of the \citet{Taylor09} catalog to Faraday thick structures is not well defined, as they had data at only two frequencies, but probably includes widths in Faraday depth up to approximately 70 \radu\ (for a Burn slab, \citealt{Burn66}, this is the Faraday thickness of the first null at 1.4 GHz). With the fractional polarizations at each frequency, we can compute the  fraction of the polarized emission in the narrow components by taking the ratio of our 150 MHz fractional polarization (which will contain only emission narrower than $\approx$1 \radu) to the \citet{Taylor09} 1.4 GHz fractional polarization (which will contain all polarization features narrower than approximately 70 \radu). This gives an upper limit to the fraction of polarized emission in narrow components, because the 1.4 GHz data will not contain all the Faraday-thick emission.

We plotted the ratio of the polarization fractions at 150 MHz and at 1.4 GHz in Fig.~\ref{fig:fracpol} (bottom panel): this figure shows that only a small fraction of sources (5 sources of 51) have more than half their polarized emission in narrow Faraday depth components. More than half of sources present in both catalogs have less than 10\% of their polarized emission (as measured at 1.4 GHz) in narrow Faraday depth components. This estimate does not include sources with no narrow components (which would not be detected in our catalog) or sources dominated by components sufficiently broad to also be depolarized at 1.4 GHz; such broad components have been observed in a few sources \citep{Anderson16}. It is possible that a few of the sources may be partially resolved at 1.4 GHz but not in our 150 MHz data, which would cause each measurement to be probing Faraday depth spectra integrated over slightly different volumes, but we expect this to affect only a small fraction of the source population being investigated and should not change our result.

From this sample, it is not clear whether the distribution in Fig.~\ref{fig:fracpol} represents a single population with a tail of narrow-component-dominated sources, or distinct populations of sources. For example, pulsars are expected, naively (i.e., without considering the frequency-dependent polarization properties of pulsar emission mechanisms), to be almost perfectly Faraday thin (we were not able to find compatible 1.4 GHz polarization data for the pulsar we observed, and could not confirm that with our data). Galaxies, both edge-on and face-on, are expected to have a high degree of Faraday complexity and correspondingly strong wavelength-dependent depolarization. Further investigation into the other properties of these sources will be necessary to determine if there is a common origin to these narrow-component dominated sources.

\subsection{Source identification and classification}

We have inspected all of the polarized sources in the full-resolution,
full-sensitivity (6\arcsec, $\sim 70$ $\mu$Jy PSF$^{-1}$) total
intensity images of the HETDEX field currently available to the LOFAR
Surveys Key Science Project (to be described by Shimwell et al.\ in
prep.), which was available for most of our detections, and also in the
1.4 GHz imaging at 5\arcsec\ resolution provided by FIRST
\citep{Becker1995}, which was available for all of them. We checked for
optical counterparts to the radio sources where possible using the
publicly available {\it WISE} band 1 and PanSTARRS I-band images.
Results are shown in Table \ref{table:ids}. We classify sources
morphologically as follows: `Compact', if the polarized source is
closest to an unresolved source in FIRST and LOFAR; `Compact double',
if the continuum source is just resolved into two components; `FR~{\sc ii}
hotspot' if the polarized source is closest to one end of a
well-resolved classical double source (which implies an angular size
of $\sim 40$\arcsec\ or greater); or `FR~{\sc ii}', if the polarized source is
associated with such a source but the position is not close to one
hotspot. A couple of other rarer morphologies are noted in the table.

The results of our inspection can be summarized as follows. Of the 89
unique polarized sources (there were two pairs of sources that were identified as belonging to different locations on the same extended radio source; the pulsar was not included in this analysis), the majority (60) are associated with
hotspots of bright classical double FR~{\sc ii}-type sources; a further 4 are
close to FR~{\sc ii}s but not unambiguously associated with the hotspots. Fifteen
are associated with unresolved sources and 7 with compact doubles.
There are a few other classifications including one unambiguous jetted
FR~{\sc i} radio galaxy but it is clear that all the compact polarized
sources in our survey, setting aside the known pulsar, are AGN; no
normal galaxies are detected although many are seen in total intensity
in the HETDEX field. Almost all the AGN are optically identified
in the data available to us, implying either that they are
quasars/blazars or (as is clearly the case for many objects) that they
are relatively nearby, $z < 0.5$ or so. The most striking source found
is the giant radio galaxy both of whose hotspots are detected in
polarization (\#45 and \#47). On the basis of a flat-spectrum core we
tentatively associate this with SDSS J123501.52+531755.0, a galaxy
with a photometric redshift of $0.47$. At this distance its angular
size of 11.5\arcmin\ would correspond to a physical size of 4.2 Mpc,
making it among the largest radio galaxies known.

\section{Discussion}\label{sec:discussion}
The catalog presented here has shown that the LOTSS data is well suited for finding polarized sources. The full LOTSS will cover the full northern sky, with a total area approximately 37 times that of this test region, and will have improved sensitivity with direction-dependent calibration, so we can expect the total polarized source count for this region to be of order 3400 sources. However, our catalog is almost certainly incomplete as described previously, so this estimate represents a conservative lower limit.

The polarization processing of this test region has highlighted a number of aspects in which the pipeline presented here could be improved. Here we summarize those aspects that we think would make the greatest improvement.

The single most limiting factor is the presence of the instrumental leakage. This puts a strong peak in the Faraday spectrum near zero, at the negative of the ionospheric leakage correction. In the \citet{Taylor09} catalog, 11\% of the sources in this region had measured RMs between $-4$ and +2 \radu, where we would expect them to be excluded in our analysis. While RM-CLEAN can strongly reduce the sidelobes from the leakage peak, some residual contamination is still present (which we suspect is due to time-variability in the ionospheric correction), concentrated around 0 \radu \ but extending all the way through the Faraday spectrum (this is reflected in the higher on-source noise). The ideal solution to this problem is to calibrate the visibilities for the polarization leakage, in the form of the so-called `d-terms' in the interferometry measurement equation \citep{Hamaker96}. While this has been performed using the SAGECal software, this is a very computationally expensive solution, and so a more efficient method is desirable. A better beam model would also contribute significantly to reducing this problem, and the development of such a model would dramatically improve the prospects for polarization studies.


Another change that could significantly reduce the number of false candidates would be to alter the imaging parameters to minimize the presence of the diffuse polarized emission. The parameters we used were optimized to maximize the sensitivity to diffuse polarized emission, which had the side-effect of making it more difficult to differentiate point sources from the diffuse foreground. Removing the short baselines, which are dominated by diffuse emission, can significantly reduce the diffuse flux present in the images \citep[e.g.,][]{Schnitzeler2009}.  The use of the polarized CLEAN algorithm \citep{Pratley2016} may also reduce the presence of artifacts in the Faraday depth cubes.

Also, it should be possible to re-image at much higher resolution, to approximately 15--30\arcsec\ using the data with only direction-independent calibration and 5--10\arcsec\ with direction-dependent calibration. Improving the resolution significantly should have three strongly beneficial effects. 
First, the polarized emission of the sources will have a much higher contrast against the diffuse polarized emission, which will make it significantly easier to identify faint point sources at the same Faraday depths as the diffuse emission. Second, we expect a population of polarized sources with angular sizes between 4 arcminutes and 20 arcseconds \citep[e.g.][]{Orru15}; these sources likely suffer from significant beam depolarization due to different polarization angles across the source being averaged together. At higher resolution, gradients in the polarization may become resolved, removing part of the depolarization and resulting in a stronger polarized signal \citep{Sokoloff98, Schnitzeler2015b}. Third, if the instrumental leakage cannot be removed (or is removed only incompletely), higher resolution will increase the effect of defocusing of the leakage in the image plane (the instrumental leakage introduces an uncalibrated interferometric phase shift, resulting in the polarization leakage having phase errors), reducing the amount of on-source leakage in the Faraday spectrum. Additionally, improved resolution will also make it easier to identify counterparts in other data (e.g., redshift surveys).

Higher resolution does introduce problems, particularly in the form of computational requirements for processing and storage of data products.  At a resolution of 20\arcsec, and comparable Faraday depth sampling to this paper, Faraday depth cubes covering the full LOTSS region will require approximately 500 TB of data storage. This will increase significantly if the Faraday depth range being probed is increased (which will be necessary closer to the Galactic plane). It may be desirable to consider a strategy that does not require full-field Faraday depth cubes, such as using the Faraday moments technique \citep{Farnes2018}, which identifies polarized source candidates in the frequency domain, or making Faraday depth cubes only around known Stokes $I$ sources (e.g., Neld et al, submitted).

To explore the data as deeply as possible, it would be advantageous to use a Stokes $I$ source catalog made from the same observations as the polarization data. At the time of this work, such a catalog was not yet available, prompting our use of the TGSS-ADR1 catalog instead. This has the limitation that the TGSS-ADR1 catalog does not go as deep as our observations; during the manual inspection of the Faraday depth cubes a few possible polarized sources were found without TGSS-ADR1 counterparts. These were found to have Stokes $I$ counterparts in the LOTSS data, but below the flux limits of the TGSS-ADR1. Using the same data for both the Stokes $I$ and polarization catalogs would ensure that the sensitivity limits are more closely matched.

If the diffuse foreground and instrumental leakage can be reduced, the number of false candidates should drop significantly and it should be possible to relax the pass/fail criteria on the source fit, or perhaps to even skip the source fitting step entirely. This would be very useful in retaining sources that deviate slightly from ideal point sources. These would include all sources that are resolved, which could be a significant population especially if the resolution is improved. While LOFAR observations are not sensitive to resolved structures in Faraday depth, it is possible to have multiple peaks in a single source. Several of these were observed in the manual inspection of the data, but none passed the pass/fail criteria. We suspect this is due to the presence of multiple peaks adjacent in Faraday depth affecting the Gaussian fit, perhaps causing it to create a fit broad enough in Faraday depth to cover both peaks, and then failing the $\sigma_\phi$ test. Being able to identify multiple-peaked sources would be an important improvement, as it would allow more detailed investigation of the Faraday depth structure of these sources.

While the processing of each observation independently allows us to compare the independent detections of sources and gives us a useful tool to verify our pipeline, it does not allow us to use the full sensitivity of the survey. By creating image mosaics from multiple observations, we could decrease the noise in the regions between pointing centers, enabling us to detect faint sources over a larger area.


\section{Summary and proposed future analysis}\label{sec:summary}
We developed a data analysis pipeline that automatically corrects the data for ionospheric Faraday rotation, identifies candidate polarized sources, and removes candidates that are due to instrumental leakage or emission from the Galactic foreground. We applied this pipeline to 60 observations from the LOFAR Two-Meter Sky Survey (LOTSS) HETDEX Spring Field region, covering a region of 570 square degrees, and identified 92 polarized radio sources (including one pulsar). This is the largest catalog of polarized sources at such a low frequency.

Our pipeline also incorporated a new error analysis, based on Monte Carlo simulation of noise with the same correlation properties as the true noise in the observations. While this method had some flaws that could not be resolved in the context of this project, we established its effectiveness by using independent measurements of the same sources. We found that the resulting error estimates produced the expected statistical behaviour. Using the same multiple independent measurements, we have shown that the systematic uncertainty introduced by the ionospheric Faraday rotation correction is closer to 0.05 \radu, which is smaller than previously predicted \citep{Sotomayor13}.

We compared our Faraday depth measurements against those observed for the same sources at 1.4 GHz in the \citet{Taylor09} catalog, finding 51 sources detected in both catalogs. There is a general correspondence between the measured Faraday depths in both catalogs, but with a large scatter that seems to be larger than can be explained by errors alone. This suggests that Faraday depth structure wider than 1 \radu \ is likely present in many of these sources, which contributes to the observed polarization at 1.4 GHz but is depolarized at 150 MHz.

We have also compared our measured polarization fractions to those at 1.4 GHz, to investigate the wavelength dependent depolarization, which we interpret in terms of narrow (non-depolarizing) and broad (depolarizing) Faraday depth components. We found that most sources are strongly depolarized in our 150 MHz data compared to the 1.4 GHz data, but some sources (approximately 10\% of the sources present in both our catalog and that of \citet{Taylor09}) show very little depolarization. Further investigation into these objects will be useful in determining if there is a distinct population of sources where narrow Faraday depth components dominate. Higher frequency, broadband observations will also be very helpful in characterizing the broad Faraday depth contributions to the polarization.

We have found a source density of 1 source per 6.2 square degrees, at 4\farcm3 resolution, but have not calculated the (polarized) flux-dependent source counts. Determining such source counts would require determinations of the incompleteness of our sample, which are not possible from the current data.
Such source counts are useful for studying the evolution of populations of radio sources; polarized source counts probe the evolution of magnetic fields. Comparing low-frequency source counts to higher-frequency source counts \citep[e.g.][]{Stil14} could provide information on the evolution of depolarization in radio sources and in turn information on the evolution of the magnetic fields.

There may be unknown pulsars among our polarized sources. Pulsars are typically highly polarized, with steep radio spectra, and are guaranteed to be point sources. Our catalog could be useful for identifying possible pulsar candidates, by cross-checking with other catalogs to obtain the spectral index and angular size of our sources. Such serendipitous pulsar discoveries have occurred with other polarization observations \citep[e.g., ][]{Navarro1995, Jelic15}. Pulsars may also have distinct properties in terms of the Faraday thin fraction, giving us an additional criterion on which to select pulsar candidates, but a larger sample is needed to check this possibility.

None of the sources in the catalog were observed to have multiple Faraday depth components (with the exception of a few resolved double-lobed sources, where each lobe was treated as a separate source). A few such sources were identified with manual inspection of the Faraday spectra, but at too low signal-to-noise ratio to be included in our catalog. Since there is a possibility that our selection criteria may bias us against detecting sources with narrowly separated Faraday depth components, we have chosen not to analyze the significance of this result.

We have shown that the LOTSS data is well suited for finding low-frequency polarized sources. The full survey will cover the entire sky above declination 0\degr; if we assume the polarized source density is the same as what we have found, we can expect to find approximately 3400 polarized sources in the full survey area. If our polarized pipeline can be improved, particularly by removing the instrumental leakage and by using the full resolution of the data, the surface density of detectable sources should rise significantly. A more robust pipeline with the ability to classify sources with multiple peaks would also be important for studying the presence of Faraday complexity in these sources. We conclude that a full polarization processing of LOTSS would be very useful in advancing the study of magnetism in distant radio sources.

\begin{acknowledgements}
This work is part of the research programme 639.042.915, which is (partly) financed by the Netherlands Organisation for Scientific Research (NWO). PNB is grateful for support from the UK STFC via grant ST/M001229/1. GJW gratefully thanks The Leverhulme Trust for the support of a Fellowship. \\

LOFAR, the Low Frequency Array designed and constructed by ASTRON, has facilities in several countries, that are owned by various parties (each with their own funding sources), and that are collectively operated by the International LOFAR Telescope (ILT) foundation under a joint scientific policy.\\

This research used ionospheric TEC maps produced by the Centre for Orbital Determination in Europe (CODE, http://aiuws.unibe.ch/ionosphere/).

This research made extensive use of Astropy, a community-developed core Python package for Astronomy \citep{Astropy}; SciPy \citep{Scipy}; NumPy \citep{Numpy}; IPython \citep{Ipython}; matplotlib \citep{Matplotlib}; and the Common Astronomy Software Applications \citep[CASA,][]{CASA}.
\end{acknowledgements}

\bibliographystyle{aa} 
\bibliography{References} 

\begin{thebibliography}{54}
\expandafter\ifx\csname natexlab\endcsname\relax\def\natexlab#1{#1}\fi

\bibitem[{{Anderson} {et~al.}(2016){Anderson}, {Gaensler}, \&
  {Feain}}]{Anderson16}
{Anderson}, C.~S., {Gaensler}, B.~M., \& {Feain}, I.~J. 2016, \apj, 825, 59

\bibitem[{{Arshakian} {et~al.}(2009){Arshakian}, {Beck}, {Krause}, \&
  {Sokoloff}}]{Arshakian2009}
{Arshakian}, T.~G., {Beck}, R., {Krause}, M., \& {Sokoloff}, D. 2009, \aap,
  494, 21

\bibitem[{{Astropy Collaboration} {et~al.}(2013){Astropy Collaboration},
  {Robitaille}, {Tollerud}, {Greenfield}, {Droettboom}, {Bray}, {Aldcroft},
  {Davis}, {Ginsburg}, {Price-Whelan}, {Kerzendorf}, {Conley}, {Crighton},
  {Barbary}, {Muna}, {Ferguson}, {Grollier}, {Parikh}, {Nair}, {Unther},
  {Deil}, {Woillez}, {Conseil}, {Kramer}, {Turner}, {Singer}, {Fox}, {Weaver},
  {Zabalza}, {Edwards}, {Azalee Bostroem}, {Burke}, {Casey}, {Crawford},
  {Dencheva}, {Ely}, {Jenness}, {Labrie}, {Lim}, {Pierfederici}, {Pontzen},
  {Ptak}, {Refsdal}, {Servillat}, \& {Streicher}}]{Astropy}
{Astropy Collaboration}, {Robitaille}, T.~P., {Tollerud}, E.~J., {et~al.} 2013,
  \aap, 558, A33

\bibitem[{{Beck} {et~al.}(2003){Beck}, {Shukurov}, {Sokoloff}, \&
  {Wielebinski}}]{Beck2003}
{Beck}, R., {Shukurov}, A., {Sokoloff}, D., \& {Wielebinski}, R. 2003, \aap,
  411, 99

\bibitem[{{Becker} {et~al.}(1995){Becker}, {White}, \& {Helfand}}]{Becker1995}
{Becker}, R.~H., {White}, R.~L., \& {Helfand}, D.~J. 1995, \apj, 450, 559

\bibitem[{{Bernardi} {et~al.}(2013){Bernardi}, {Greenhill}, {Mitchell}, {Ord},
  {Hazelton}, {Gaensler}, {de Oliveira-Costa}, {Morales}, {Udaya Shankar},
  {Subrahmanyan}, {Wayth}, {Lenc}, {Williams}, {Arcus}, {Arora}, {Barnes},
  {Bowman}, {Briggs}, {Bunton}, {Cappallo}, {Corey}, {Deshpande}, {deSouza},
  {Emrich}, {Goeke}, {Herne}, {Hewitt}, {Johnston-Hollitt}, {Kaplan}, {Kasper},
  {Kincaid}, {Koenig}, {Kratzenberg}, {Lonsdale}, {Lynch}, {McWhirter},
  {Morgan}, {Oberoi}, {Pathikulangara}, {Prabu}, {Remillard}, {Rogers},
  {Roshi}, {Salah}, {Sault}, {Srivani}, {Stevens}, {Tingay}, {Waterson},
  {Webster}, {Whitney}, {Williams}, \& {Wyithe}}]{Bernardi13}
{Bernardi}, G., {Greenhill}, L.~J., {Mitchell}, D.~A., {et~al.} 2013, \apj,
  771, 105

\bibitem[{{Bilous} {et~al.}(2016){Bilous}, {Kondratiev}, {Kramer}, {Keane},
  {Hessels}, {Stappers}, {Malofeev}, {Sobey}, {Breton}, {Cooper}, {Falcke},
  {Karastergiou}, {Michilli}, {Os{\l}owski}, {Sanidas}, {ter Veen}, {van
  Leeuwen}, {Verbiest}, {Weltevrede}, {Zarka}, {Grie{\ss}meier}, {Serylak},
  {Bell}, {Broderick}, {Eisl{\"o}ffel}, {Markoff}, \& {Rowlinson}}]{Bilous16}
{Bilous}, A.~V., {Kondratiev}, V.~I., {Kramer}, M., {et~al.} 2016, \aap, 591,
  A134

\bibitem[{{Blasi}(2013)}]{Blasi2013}
{Blasi}, P. 2013, \aapr, 21, 70

\bibitem[{{Brentjens} \& {de Bruyn}(2005)}]{Brentjens05}
{Brentjens}, M.~A. \& {de Bruyn}, A.~G. 2005, \aap, 441, 1217

\bibitem[{{Burn}(1966)}]{Burn66}
{Burn}, B.~J. 1966, \mnras, 133, 67

\bibitem[{{Cordes} \& {Lazio}(2002)}]{NE2001}
{Cordes}, J.~M. \& {Lazio}, T.~J.~W. 2002, ArXiv Astrophysics e-prints
  [\eprint{astro-ph/0207156}]

\bibitem[{{Falceta-Gon{\c c}alves} {et~al.}(2014){Falceta-Gon{\c c}alves},
  {Kowal}, {Falgarone}, \& {Chian}}]{Falceta2014}
{Falceta-Gon{\c c}alves}, D., {Kowal}, G., {Falgarone}, E., \& {Chian},
  A.~C.-L. 2014, Nonlinear Processes in Geophysics, 21, 587

\bibitem[{{Farnes} {et~al.}(2018){Farnes}, {Heald}, {Junklewitz}, {Mulcahy},
  {Haverkorn}, {Van Eck}, {Riseley}, {Brentjens}, {Horellou}, {Vacca}, {Jones},
  {Horneffer}, \& {Paladino}}]{Farnes2018}
{Farnes}, J.~S., {Heald}, G., {Junklewitz}, H., {et~al.} 2018, \mnras, 474,
  3280

\bibitem[{{Farnes} {et~al.}(2014){Farnes}, {O'Sullivan}, {Corrigan}, \&
  {Gaensler}}]{Farnes14}
{Farnes}, J.~S., {O'Sullivan}, S.~P., {Corrigan}, M.~E., \& {Gaensler}, B.~M.
  2014, \apj, 795, 63

\bibitem[{{Force} {et~al.}(2015){Force}, {Demorest}, \& {Rankin}}]{Force2015}
{Force}, M.~M., {Demorest}, P., \& {Rankin}, J.~M. 2015, \mnras, 453, 4485

\bibitem[{Hales {et~al.}(2012)Hales, Gaensler, Norris, \&
  Middelberg}]{Hales2012}
Hales, C.~A., Gaensler, B.~M., Norris, R.~P., \& Middelberg, E. 2012, Monthly
  Notices of the Royal Astronomical Society, 424, 2160

\bibitem[{Hamaker {et~al.}(1996)Hamaker, Bregman, \& Sault}]{Hamaker96}
Hamaker, J.~P., Bregman, J.~D., \& Sault, R.~J. 1996, Astronomy and
  Astrophysics Supplement Series, 117, 137

\bibitem[{{Han} {et~al.}(1998){Han}, {Beck}, \& {Berkhuijsen}}]{Han1998}
{Han}, J.~L., {Beck}, R., \& {Berkhuijsen}, E.~M. 1998, \aap, 335, 1117

\bibitem[{{Haverkorn} {et~al.}(2008){Haverkorn}, {Brown}, {Gaensler}, \&
  {McClure-Griffiths}}]{Haverkorn08}
{Haverkorn}, M., {Brown}, J.~C., {Gaensler}, B.~M., \& {McClure-Griffiths},
  N.~M. 2008, \apj, 680, 362

\bibitem[{{Hill} {et~al.}(2008){Hill}, {Gebhardt}, {Komatsu}, {Drory},
  {MacQueen}, {Adams}, {Blanc}, {Koehler}, {Rafal}, {Roth}, {Kelz}, {Gronwall},
  {Ciardullo}, \& {Schneider}}]{Hill08}
{Hill}, G.~J., {Gebhardt}, K., {Komatsu}, E., {et~al.} 2008, in Astronomical
  Society of the Pacific Conference Series, Vol. 399, Panoramic Views of Galaxy
  Formation and Evolution, ed. T.~{Kodama}, T.~{Yamada}, \& K.~{Aoki}, 115

\bibitem[{Hunter(2007)}]{Matplotlib}
Hunter, J.~D. 2007, Computing In Science \& Engineering, 9, 90

\bibitem[{{Intema} {et~al.}(2017){Intema}, {Jagannathan}, {Mooley}, \&
  {Frail}}]{Intema2017}
{Intema}, H.~T., {Jagannathan}, P., {Mooley}, K.~P., \& {Frail}, D.~A. 2017,
  \aap, 598, A78

\bibitem[{Jansson \& Farrar(2012)}]{Jansson2012}
Jansson, R. \& Farrar, G.~R. 2012, The Astrophysical Journal, 757, 14

\bibitem[{{Jeli{\'c}} {et~al.}(2015){Jeli{\'c}}, {de Bruyn}, {Pandey},
  {Mevius}, {Haverkorn}, {Brentjens}, {Koopmans}, {Zaroubi}, {Abdalla}, {Asad},
  {Bus}, {Chapman}, {Ciardi}, {Fernandez}, {Ghosh}, {Harker}, {Iliev},
  {Jensen}, {Kazemi}, {Mellema}, {Offringa}, {Patil}, {Vedantham}, \&
  {Yatawatta}}]{Jelic15}
{Jeli{\'c}}, V., {de Bruyn}, A.~G., {Pandey}, V.~N., {et~al.} 2015, \aap, 583,
  A137

\bibitem[{Jones {et~al.}(2001--)Jones, Oliphant, Peterson, {et~al.}}]{Scipy}
Jones, E., Oliphant, T., Peterson, P., {et~al.} 2001--, {SciPy}: Open source
  scientific tools for {Python}, [Online: www.scipy.org]

\bibitem[{{Lamee} {et~al.}(2016){Lamee}, {Rudnick}, {Farnes}, {Carretti},
  {Gaensler}, {Haverkorn}, \& {Poppi}}]{Lamee2016}
{Lamee}, M., {Rudnick}, L., {Farnes}, J.~S., {et~al.} 2016, \apj, 829, 5

\bibitem[{{Lewis} \& {Bate}(2017)}]{Lewis17}
{Lewis}, B.~T. \& {Bate}, M.~R. 2017, ArXiv e-prints
  [\eprint[arXiv]{1701.08741}]

\bibitem[{Macquart {et~al.}(2012)Macquart, Ekers, Feain, \&
  Johnston-Hollitt}]{Macquart2012}
Macquart, J.-P., Ekers, R.~D., Feain, I., \& Johnston-Hollitt, M. 2012, The
  Astrophysical Journal, 750, 139

\bibitem[{{Manchester} {et~al.}(2005){Manchester}, {Hobbs}, {Teoh}, \&
  {Hobbs}}]{ATNF}
{Manchester}, R.~N., {Hobbs}, G.~B., {Teoh}, A., \& {Hobbs}, M. 2005, \aj, 129,
  1993

\bibitem[{{McMullin} {et~al.}(2007){McMullin}, {Waters}, {Schiebel}, {Young},
  \& {Golap}}]{CASA}
{McMullin}, J.~P., {Waters}, B., {Schiebel}, D., {Young}, W., \& {Golap}, K.
  2007, in Astronomical Society of the Pacific Conference Series, Vol. 376,
  Astronomical Data Analysis Software and Systems XVI, ed. R.~A. {Shaw},
  F.~{Hill}, \& D.~J. {Bell}, 127

\bibitem[{{Mulcahy} {et~al.}(2014){Mulcahy}, {Horneffer}, {Beck}, {Heald},
  {Fletcher}, {Scaife}, {Adebahr}, {Anderson}, {Bonafede}, {Br{\"u}ggen},
  {Brunetti}, {Chy{\.z}y}, {Conway}, {Dettmar}, {En{\ss}lin}, {Haverkorn},
  {Horellou}, {Iacobelli}, {Israel}, {Junklewitz}, {Jurusik}, {K{\"o}hler},
  {Kuniyoshi}, {Orr{\'u}}, {Paladino}, {Pizzo}, {Reich}, \&
  {R{\"o}ttgering}}]{Mulcahy14}
{Mulcahy}, D.~D., {Horneffer}, A., {Beck}, R., {et~al.} 2014, \aap, 568, A74

\bibitem[{{Navarro} {et~al.}(1995){Navarro}, {de Bruyn}, {Frail}, {Kulkarni},
  \& {Lyne}}]{Navarro1995}
{Navarro}, J., {de Bruyn}, A.~G., {Frail}, D.~A., {Kulkarni}, S.~R., \& {Lyne},
  A.~G. 1995, \apjl, 455, L55

\bibitem[{{Orr{\`u}} {et~al.}(2015){Orr{\`u}}, {van Velzen}, {Pizzo},
  {Yatawatta}, {Paladino}, {Iacobelli}, {Murgia}, {Falcke}, {Morganti}, {de
  Bruyn}, {Ferrari}, {Anderson}, {Bonafede}, {Mulcahy}, {Asgekar}, {Avruch},
  {Beck}, {Bell}, {van Bemmel}, {Bentum}, {Bernardi}, {Best}, {Breitling},
  {Broderick}, {Br{\"u}ggen}, {Butcher}, {Ciardi}, {Conway}, {Corstanje}, {de
  Geus}, {Deller}, {Duscha}, {Eisl{\"o}ffel}, {Engels}, {Frieswijk}, {Garrett},
  {Grie{\ss}meier}, {Gunst}, {Hamaker}, {Heald}, {Hoeft}, {van der Horst},
  {Intema}, {Juette}, {Kohler}, {Kondratiev}, {Kuniyoshi}, {Kuper}, {Loose},
  {Maat}, {Mann}, {Markoff}, {McFadden}, {McKay-Bukowski}, {Miley}, {Moldon},
  {Molenaar}, {Munk}, {Nelles}, {Paas}, {Pandey-Pommier}, {Pandey}, {Pietka},
  {Polatidis}, {Reich}, {R{\"o}ttgering}, {Rowlinson}, {Scaife},
  {Schoenmakers}, {Schwarz}, {Serylak}, {Shulevski}, {Smirnov}, {Steinmetz},
  {Stewart}, {Swinbank}, {Tagger}, {Tasse}, {Thoudam}, {Toribio}, {Vermeulen},
  {Vocks}, {van Weeren}, {Wijers}, {Wise}, \& {Wucknitz}}]{Orru15}
{Orr{\`u}}, E., {van Velzen}, S., {Pizzo}, R.~F., {et~al.} 2015, \aap, 584,
  A112

\bibitem[{P\'erez \& Granger(2007)}]{Ipython}
P\'erez, F. \& Granger, B.~E. 2007, Computing in Science and Engineering, 9, 21

\bibitem[{{Pratley} \& {Johnston-Hollitt}(2016)}]{Pratley2016}
{Pratley}, L. \& {Johnston-Hollitt}, M. 2016, \mnras, 462, 3483

\bibitem[{{Schnitzeler} {et~al.}(2015){Schnitzeler}, {Banfield}, \&
  {Lee}}]{Schnitzeler2015b}
{Schnitzeler}, D.~H.~F.~M., {Banfield}, J.~K., \& {Lee}, K.~J. 2015, \mnras,
  450, 3579

\bibitem[{{Schnitzeler} {et~al.}(2009){Schnitzeler}, {Katgert}, \& {de
  Bruyn}}]{Schnitzeler2009}
{Schnitzeler}, D.~H.~F.~M., {Katgert}, P., \& {de Bruyn}, A.~G. 2009, \aap,
  494, 611

\bibitem[{{Shimwell} {et~al.}(2017){Shimwell}, {R{\"o}ttgering}, {Best},
  {Williams}, {Dijkema}, {de Gasperin}, {Hardcastle}, {Heald}, {Hoang},
  {Horneffer}, {Intema}, {Mahony}, {Mandal}, {Mechev}, {Morabito}, {Oonk},
  {Rafferty}, {Retana-Montenegro}, {Sabater}, {Tasse}, {van Weeren},
  {Br{\"u}ggen}, {Brunetti}, {Chy{\.z}y}, {Conway}, {Haverkorn}, {Jackson},
  {Jarvis}, {McKean}, {Miley}, {Morganti}, {White}, {Wise}, {van Bemmel},
  {Beck}, {Brienza}, {Bonafede}, {Calistro Rivera}, {Cassano}, {Clarke},
  {Cseh}, {Deller}, {Drabent}, {van Driel}, {Engels}, {Falcke}, {Ferrari},
  {Fr{\"o}hlich}, {Garrett}, {Harwood}, {Heesen}, {Hoeft}, {Horellou},
  {Israel}, {Kapi{\'n}ska}, {Kunert-Bajraszewska}, {McKay}, {Mohan},
  {Orr{\'u}}, {Pizzo}, {Prandoni}, {Schwarz}, {Shulevski}, {Sipior}, {Smith},
  {Sridhar}, {Steinmetz}, {Stroe}, {Varenius}, {van der Werf}, {Zensus}, \&
  {Zwart}}]{Shimwell17}
{Shimwell}, T.~W., {R{\"o}ttgering}, H.~J.~A., {Best}, P.~N., {et~al.} 2017,
  \aap, 598, A104

\bibitem[{{Simmons} \& {Stewart}(1985)}]{Simmons85}
{Simmons}, J.~F.~L. \& {Stewart}, B.~G. 1985, \aap, 142, 100

\bibitem[{{Sokoloff} {et~al.}(1998){Sokoloff}, {Bykov}, {Shukurov},
  {Berkhuijsen}, {Beck}, \& {Poezd}}]{Sokoloff98}
{Sokoloff}, D.~D., {Bykov}, A.~A., {Shukurov}, A., {et~al.} 1998, \mnras, 299,
  189

\bibitem[{{Sotomayor-Beltran} {et~al.}(2013){Sotomayor-Beltran}, {Sobey},
  {Hessels}, {de Bruyn}, {Noutsos}, {Alexov}, {Anderson}, {Asgekar}, {Avruch},
  {Beck}, {Bell}, {Bell}, {Bentum}, {Bernardi}, {Best}, {Birzan}, {Bonafede},
  {Breitling}, {Broderick}, {Brouw}, {Br{\"u}ggen}, {Ciardi}, {de Gasperin},
  {Dettmar}, {van Duin}, {Duscha}, {Eisl{\"o}ffel}, {Falcke}, {Fallows},
  {Fender}, {Ferrari}, {Frieswijk}, {Garrett}, {Grie{\ss}meier}, {Grit},
  {Gunst}, {Hassall}, {Heald}, {Hoeft}, {Horneffer}, {Iacobelli}, {Juette},
  {Karastergiou}, {Keane}, {Kohler}, {Kramer}, {Kondratiev}, {Koopmans},
  {Kuniyoshi}, {Kuper}, {van Leeuwen}, {Maat}, {Macario}, {Markoff}, {McKean},
  {Mulcahy}, {Munk}, {Orru}, {Paas}, {Pandey-Pommier}, {Pilia}, {Pizzo},
  {Polatidis}, {Reich}, {R{\"o}ttgering}, {Serylak}, {Sluman}, {Stappers},
  {Tagger}, {Tang}, {Tasse}, {ter Veen}, {Vermeulen}, {van Weeren}, {Wijers},
  {Wijnholds}, {Wise}, {Wucknitz}, {Yatawatta}, \& {Zarka}}]{Sotomayor13}
{Sotomayor-Beltran}, C., {Sobey}, C., {Hessels}, J.~W.~T., {et~al.} 2013, \aap,
  552, A58

\bibitem[{{Stil} {et~al.}(2014){Stil}, {Keller}, {George}, \&
  {Taylor}}]{Stil14}
{Stil}, J.~M., {Keller}, B.~W., {George}, S.~J., \& {Taylor}, A.~R. 2014, \apj,
  787, 99

\bibitem[{{Stil} {et~al.}(2011){Stil}, {Taylor}, \& {Sunstrum}}]{Stil11}
{Stil}, J.~M., {Taylor}, A.~R., \& {Sunstrum}, C. 2011, \apj, 726, 4

\bibitem[{{Tasse} {et~al.}(2013){Tasse}, {van der Tol}, {van Zwieten}, {van
  Diepen}, \& {Bhatnagar}}]{Tasse13}
{Tasse}, C., {van der Tol}, S., {van Zwieten}, J., {van Diepen}, G., \&
  {Bhatnagar}, S. 2013, \aap, 553, A105

\bibitem[{{Taylor} {et~al.}(2009){Taylor}, {Stil}, \& {Sunstrum}}]{Taylor09}
{Taylor}, A.~R., {Stil}, J.~M., \& {Sunstrum}, C. 2009, \apj, 702, 1230

\bibitem[{{Tingay} {et~al.}(2013){Tingay}, {Goeke}, {Bowman}, {Emrich}, {Ord},
  {Mitchell}, {Morales}, {Booler}, {Crosse}, {Wayth}, {Lonsdale}, {Tremblay},
  {Pallot}, {Colegate}, {Wicenec}, {Kudryavtseva}, {Arcus}, {Barnes},
  {Bernardi}, {Briggs}, {Burns}, {Bunton}, {Cappallo}, {Corey}, {Deshpande},
  {Desouza}, {Gaensler}, {Greenhill}, {Hall}, {Hazelton}, {Herne}, {Hewitt},
  {Johnston-Hollitt}, {Kaplan}, {Kasper}, {Kincaid}, {Koenig}, {Kratzenberg},
  {Lynch}, {Mckinley}, {Mcwhirter}, {Morgan}, {Oberoi}, {Pathikulangara},
  {Prabu}, {Remillard}, {Rogers}, {Roshi}, {Salah}, {Sault}, {Udaya-Shankar},
  {Schlagenhaufer}, {Srivani}, {Stevens}, {Subrahmanyan}, {Waterson},
  {Webster}, {Whitney}, {Williams}, {Williams}, \& {Wyithe}}]{Tingay2013}
{Tingay}, S.~J., {Goeke}, R., {Bowman}, J.~D., {et~al.} 2013, \pasa, 30, e007

\bibitem[{{Vacca} {et~al.}(2016){Vacca}, {Oppermann}, {En{\ss}lin}, {Jasche},
  {Selig}, {Greiner}, {Junklewitz}, {Reinecke}, {Br{\"u}ggen}, {Carretti},
  {Feretti}, {Ferrari}, {Hales}, {Horellou}, {Ideguchi}, {Johnston-Hollitt},
  {Pizzo}, {R{\"o}ttgering}, {Shimwell}, \& {Takahashi}}]{Vacca2016}
{Vacca}, V., {Oppermann}, N., {En{\ss}lin}, T., {et~al.} 2016, \aap, 591, A13

\bibitem[{van~der Walt {et~al.}(2011)van~der Walt, Colbert, \&
  Varoquaux}]{Numpy}
van~der Walt, S., Colbert, S.~C., \& Varoquaux, G. 2011, Computing in Science
  \& Engineering, 13, 22

\bibitem[{{Van Eck}(2017)}]{MyThesis}
{Van Eck}, C.~L. 2017, PhD thesis, Radboud University Nijmegen

\bibitem[{{Van Eck} {et~al.}(2011){Van Eck}, {Brown}, {Stil}, {Rae}, {Mao},
  {Gaensler}, {Shukurov}, {Taylor}, {Haverkorn}, {Kronberg}, \&
  {McClure-Griffiths}}]{VanEck11}
{Van Eck}, C.~L., {Brown}, J.~C., {Stil}, J.~M., {et~al.} 2011, \apj, 728, 97

\bibitem[{{Van Eck} {et~al.}(2017){Van Eck}, {Haverkorn}, {Alves}, {Beck}, {de
  Bruyn}, {En{\ss}lin}, {Farnes}, {Ferri{\`e}re}, {Heald}, {Horellou},
  {Horneffer}, {Iacobelli}, {Jeli{\'c}}, {Mart{\'{\i}}-Vidal}, {Mulcahy},
  {Reich}, {R{\"o}ttgering}, {Scaife}, {Schnitzeler}, {Sobey}, \&
  {Sridhar}}]{VanEck17}
{Van Eck}, C.~L., {Haverkorn}, M., {Alves}, M.~I.~R., {et~al.} 2017, \aap, 597,
  A98

\bibitem[{{van Haarlem} {et~al.}(2013){van Haarlem}, {Wise}, {Gunst}, {Heald},
  {McKean}, {Hessels}, {de Bruyn}, {Nijboer}, {Swinbank}, {Fallows},
  {Brentjens}, {Nelles}, {Beck}, {Falcke}, {Fender}, {H{\"o}randel},
  {Koopmans}, {Mann}, {Miley}, {R{\"o}ttgering}, {Stappers}, {Wijers},
  {Zaroubi}, {van den Akker}, {Alexov}, {Anderson}, {Anderson}, {van Ardenne},
  {Arts}, {Asgekar}, {Avruch}, {Batejat}, {B{\"a}hren}, {Bell}, {Bell}, {van
  Bemmel}, {Bennema}, {Bentum}, {Bernardi}, {Best}, {B{\^\i}rzan}, {Bonafede},
  {Boonstra}, {Braun}, {Bregman}, {Breitling}, {van de Brink}, {Broderick},
  {Broekema}, {Brouw}, {Br{\"u}ggen}, {Butcher}, {van Cappellen}, {Ciardi},
  {Coenen}, {Conway}, {Coolen}, {Corstanje}, {Damstra}, {Davies}, {Deller},
  {Dettmar}, {van Diepen}, {Dijkstra}, {Donker}, {Doorduin}, {Dromer}, {Drost},
  {van Duin}, {Eisl{\"o}ffel}, {van Enst}, {Ferrari}, {Frieswijk}, {Gankema},
  {Garrett}, {de Gasperin}, {Gerbers}, {de Geus}, {Grie{\ss}meier}, {Grit},
  {Gruppen}, {Hamaker}, {Hassall}, {Hoeft}, {Holties}, {Horneffer}, {van der
  Horst}, {van Houwelingen}, {Huijgen}, {Iacobelli}, {Intema}, {Jackson},
  {Jelic}, {de Jong}, {Juette}, {Kant}, {Karastergiou}, {Koers}, {Kollen},
  {Kondratiev}, {Kooistra}, {Koopman}, {Koster}, {Kuniyoshi}, {Kramer},
  {Kuper}, {Lambropoulos}, {Law}, {van Leeuwen}, {Lemaitre}, {Loose}, {Maat},
  {Macario}, {Markoff}, {Masters}, {McFadden}, {McKay-Bukowski}, {Meijering},
  {Meulman}, {Mevius}, {Middelberg}, {Millenaar}, {Miller-Jones}, {Mohan},
  {Mol}, {Morawietz}, {Morganti}, {Mulcahy}, {Mulder}, {Munk}, {Nieuwenhuis},
  {van Nieuwpoort}, {Noordam}, {Norden}, {Noutsos}, {Offringa}, {Olofsson},
  {Omar}, {Orr{\'u}}, {Overeem}, {Paas}, {Pandey-Pommier}, {Pandey}, {Pizzo},
  {Polatidis}, {Rafferty}, {Rawlings}, {Reich}, {de Reijer}, {Reitsma},
  {Renting}, {Riemers}, {Rol}, {Romein}, {Roosjen}, {Ruiter}, {Scaife}, {van
  der Schaaf}, {Scheers}, {Schellart}, {Schoenmakers}, {Schoonderbeek},
  {Serylak}, {Shulevski}, {Sluman}, {Smirnov}, {Sobey}, {Spreeuw}, {Steinmetz},
  {Sterks}, {Stiepel}, {Stuurwold}, {Tagger}, {Tang}, {Tasse}, {Thomas},
  {Thoudam}, {Toribio}, {van der Tol}, {Usov}, {van Veelen}, {van der Veen},
  {ter Veen}, {Verbiest}, {Vermeulen}, {Vermaas}, {Vocks}, {Vogt}, {de Vos},
  {van der Wal}, {van Weeren}, {Weggemans}, {Weltevrede}, {White}, {Wijnholds},
  {Wilhelmsson}, {Wucknitz}, {Yatawatta}, {Zarka}, {Zensus}, \& {van
  Zwieten}}]{vanHaarlem2013}
{van Haarlem}, M.~P., {Wise}, M.~W., {Gunst}, A.~W., {et~al.} 2013, \aap, 556,
  A2

\bibitem[{{Yao} {et~al.}(2017){Yao}, {Manchester}, \& {Wang}}]{YMW16}
{Yao}, J.~M., {Manchester}, R.~N., \& {Wang}, N. 2017, \apj, 835, 29

\bibitem[{{Yatawatta} {et~al.}(2008){Yatawatta}, {Zaroubi}, {de Bruyn},
  {Koopmans}, \& {Noordam}}]{Yatawatta2008}
{Yatawatta}, S., {Zaroubi}, S., {de Bruyn}, G., {Koopmans}, L., \& {Noordam},
  J. 2008, ArXiv e-prints [\eprint[arXiv]{0810.5751}]

\end{thebibliography}

\appendix
\section{Tables}

\longtab[1]{
\begin{landscape}
\begin{longtable}{rcr@{ $\pm$ }lr@{ $\pm$ }lcr@{ $\pm$ }lr@{ $\pm$ }lr@{ $\pm$ }lr@{ $\pm$ }lr@{ $\pm$ }l}
\caption{Catalog of polarized point sources \label{tab:catalog}}\\
\hline
ID & TGSS-ADR1 ID &  \multicolumn{2}{c}{RA (J2000)}& \multicolumn{2}{c}{Dec (J2000)} & N\tablefootmark{a} & \multicolumn{2}{c}{Pol. Int.}  & \multicolumn{2}{c}{$\phi$} & \multicolumn{2}{c}{$m$\tablefootmark{b}} & \multicolumn{2}{c}{1.4 GHz RM\tablefootmark{c}} & \multicolumn{2}{c}{1.4 GHz $m$\tablefootmark{c}} \\
 & & \multicolumn{2}{c}{[h m s]} & \multicolumn{2}{c}{[\degr\ \arcmin \ \arcsec]} &  & \multicolumn{2}{c}{[mJy PSF\inv]} & \multicolumn{2}{c}{[\radu]} & \multicolumn{2}{c}{[\%]} & \multicolumn{2}{c}{[\radu]} & \multicolumn{2}{c}{[\%]} \\
\hline
\endfirsthead
\caption{continued.}\\
\hline
ID & TGSS-ADR1 ID &  \multicolumn{2}{c}{RA (J2000)}& \multicolumn{2}{c}{Dec (J2000)} & N\tablefootmark{a} & \multicolumn{2}{c}{Pol. Int.}  & \multicolumn{2}{c}{$\phi$} & \multicolumn{2}{c}{$m$\tablefootmark{b}} & \multicolumn{2}{c}{1.4 GHz RM\tablefootmark{c}} & \multicolumn{2}{c}{1.4 GHz $m$\tablefootmark{c}} \\
 & & \multicolumn{2}{c}{[h m s]} & \multicolumn{2}{c}{[\degr\ \arcmin \ \arcsec]} &  & \multicolumn{2}{c}{[mJy PSF\inv]} & \multicolumn{2}{c}{[\radu]} & \multicolumn{2}{c}{[\%]} & \multicolumn{2}{c}{[\radu]} & \multicolumn{2}{c}{[\%]} \\
\hline
\endhead
\hline
\multicolumn{17}{l}{\tablefoottext{a}{Number of independent detections.} \tablefoottext{b}{Fractional polarization ($P$/$I$).} \tablefoottext{c}{From the \citet{Taylor09} catalog; sources without a match are marked with '--'.} }
\endfoot
\hline
\multicolumn{17}{l}{\tablefoottext{a}{Number of independent detections.} \tablefoottext{b}{Fractional polarization ($P$/$I$).} \tablefoottext{c}{From the \citet{Taylor09} catalog; sources without a match are marked with '--'.} }\\
 \multicolumn{17}{l}{$^{(*)}$ Pulsar B1112+50/J1115+5030}
\endlastfoot
  1 & J104221.7+530524 & 10 42 22.6&0.9 & +53 05 24.2&09.4 & 1 & 3.63&0.19 & +12.92&0.05 & 1.52&0.17 &  \multicolumn{2}{c}{--} & \multicolumn{2}{c}{--} \\
  2 & J104628.4+544944 & 10 46 30.1&0.3 & +54 49 38.1&02.6 & 1 & 5.84&0.13 & +8.94&0.03 & 2.31&0.24 & 22.1&12.7 & 5.90&0.35 \\
  3 & J104641.0+543424 & 10 46 41.1&0.2 & +54 34 26.1&01.6 & 2 & 24.66&0.31 & +8.29&0.01 & 4.90&0.50 &  7.9&11.8 & 7.72&0.62 \\
  4 & J105410.0+471441 & 10 54 12.0&1.8 & +47 14 53.9&20.5 & 1 & 0.81&0.12 & +28.24&0.14 & 0.85&0.16 &  \multicolumn{2}{c}{--} & \multicolumn{2}{c}{--} \\
  5 & J105500.7+520202 & 10 55 02.8&0.6 & +52 01 57.6&06.0 & 3 & 8.81&0.42 & +17.82&0.07 & 0.41&0.05 &  \multicolumn{2}{c}{--} & \multicolumn{2}{c}{--} \\
  6 & J105702.8+483652 & 10 57 02.9&0.1 & +48 36 42.7&01.4 & 3 & 14.99&0.14 & +16.79&0.01 & 4.93&0.50 & 15.0& 9.4 & 12.74&0.58 \\
  7 & J105706.7+532543 & 10 57 07.1&0.3 & +53 25 41.3&01.9 & 3 & 37.82&0.54 & +12.69&0.01 & 3.49&0.35 &  5.7& 4.3 & 7.10&0.14 \\
  8 & J110009.3+494022 & 11 00 10.5&0.7 & +49 40 36.8&06.7 & 1 & 1.30&0.06 & +16.34&0.06 & 2.70&0.37 &  \multicolumn{2}{c}{--} & \multicolumn{2}{c}{--} \\
  9 & J110137.7+515701 & 11 01 41.2&0.8 & +51 57 05.3&06.5 & 1 & 1.63&0.08 & +12.45&0.06 & 8.11&1.82 &  \multicolumn{2}{c}{--} & \multicolumn{2}{c}{--} \\
 10 & J110208.0+474328 & 11 02 10.3&0.7 & +47 43 32.2&07.2 & 1 & 2.30&0.10 & +23.68&0.04 & 5.51&0.81 &  \multicolumn{2}{c}{--} & \multicolumn{2}{c}{--} \\
 11 & J110249.5+531247 & 11 02 48.1&1.2 & +53 12 37.8&08.9 & 2 & 6.15&0.42 & +18.33&0.08 & 0.84&0.10 & 17.2& 8.0 & 5.61&0.22 \\
 12 & J110305.0+525940 & 11 03 04.7&1.1 & +52 59 29.0&10.1 & 1 & 1.47&0.13 & +15.10&0.14 & 0.72&0.10 & 30.4&16.5 & 6.48&0.48 \\
 13 & J110638.1+542951 & 11 06 40.7&0.7 & +54 30 02.1&05.2 & 2 & 2.55&0.09 & +13.78&0.03 & 0.63&0.07 &  \multicolumn{2}{c}{--} & \multicolumn{2}{c}{--} \\
 14 & J110941.6+531242 & 11 09 40.0&0.7 & +53 12 37.9&06.4 & 1 & 2.29&0.11 & +11.66&0.08 & 1.12&0.13 & 18.8& 8.2 & 13.58&0.61 \\
 15 & J112023.1+540427 & 11 20 23.9&0.9 & +54 04 30.0&07.4 & 3 & 4.03&0.26 & +21.07&0.08 & 0.36&0.04 & 18.7& 5.1 & 3.90&0.10 \\
 16 & J112026.4+571000 & 11 20 24.8&0.4 & +57 10 33.2&02.9 & 2 & 14.17&0.28 & +13.07&0.02 & 1.40&0.14 &  \multicolumn{2}{c}{--} & \multicolumn{2}{c}{--} \\
 17 & J112353.8+514148 & 11 23 53.9&0.8 & +51 41 42.2&07.0 & 2 & 3.14&0.14 & +7.80&0.04 & 0.27&0.03 & 18.4&12.8 & 3.49&0.20 \\
 18 & J112542.0+564224 & 11 25 45.4&1.1 & +56 42 30.2&06.4 & 1 & 3.12&0.19 & +13.83&0.06 & 0.84&0.10 & -11.8&16.2 & 7.03&0.54 \\
 19 & J112606.0+502217 & 11 26 04.7&0.4 & +50 21 57.5&05.0 & 2 & 6.08&0.19 & +6.75&0.04 & 0.57&0.06 &  5.7& 5.0 & 9.81&0.22 \\
 20 & J113756.1+471312 & 11 37 57.2&3.5 & +47 12 58.4&36.6 & 1 & 1.20&0.22 & +21.82&0.22 & 1.96&0.45 &  \multicolumn{2}{c}{--} & \multicolumn{2}{c}{--} \\
 21 & J113817.0+495023 & 11 38 15.7&0.7 & +49 50 34.1&05.1 & 1 & 1.60&0.07 & +9.09&0.06 & 0.53&0.06 & -1.5&13.1 & 6.77&0.41 \\
 22 & J115231.6+463113 & 11 52 32.1&1.4 & +46 31 00.1&20.3 & 1 & 1.32&0.17 & +16.36&0.09 & 0.58&0.10 &  \multicolumn{2}{c}{--} & \multicolumn{2}{c}{--} \\
 23 & J115316.2+480358 & 11 53 15.9&1.8 & +48 03 56.8&19.1 & 1 & 1.06&0.15 & +26.38&0.16 & 0.67&0.12 & 33.5&16.6 & 8.54&0.64 \\
 24 & J115405.8+562040 & 11 54 07.4&1.1 & +56 20 47.9&06.6 & 1 & 13.87&0.70 & +2.86&0.05 & 1.02&0.11 &  \multicolumn{2}{c}{--} & \multicolumn{2}{c}{--} \\
 25 & J115405.8+562040 & 11 54 09.9&0.4 & +56 21 08.5&02.8 & 1 & 6.84&0.14 & -3.88&0.02 & 0.50&0.05 &  \multicolumn{2}{c}{--} & \multicolumn{2}{c}{--} \\
 26 & J115420.7+452330 & 11 54 20.8&0.1 & +45 24 01.6&01.1 & 2 & 83.99&0.68 & +8.37&0.01 & 1.59&0.16 & 10.7& 5.4 & 2.10&0.06 \\
 27 & J115907.2+483831 & 11 59 12.4&1.5 & +48 38 39.1&14.6 & 1 & 1.05&0.12 & +33.92&0.12 & 0.16&0.02 & 11.8&10.0 & 7.43&0.36 \\
 28 & J120125.1+492626 & 12 01 30.0&0.4 & +49 27 17.5&03.8 & 3 & 5.98&0.18 & +22.77&0.03 & 0.50&0.05 &  \multicolumn{2}{c}{--} & \multicolumn{2}{c}{--} \\
 29 & J120607.9+521158 & 12 06 06.5&1.4 & +52 12 17.9&12.0 & 1 & 3.29&0.30 & +12.56&0.10 & 0.25&0.03 & 47.0&10.3 & 3.18&0.15 \\
 30 & J121041.5+532907 & 12 10 40.5&0.4 & +53 29 02.7&03.9 & 3 & 7.81&0.20 & +12.08&0.02 & 11.69&1.53 &  \multicolumn{2}{c}{--} & \multicolumn{2}{c}{--} \\
 31 & J121043.5+483424 & 12 10 44.0&0.6 & +48 34 30.5&05.7 & 2 & 2.49&0.09 & +11.99&0.04 & 0.36&0.04 & 13.7& 7.5 & 9.09&0.32 \\
 32 & J121158.7+545602 & 12 11 59.0&0.2 & +54 56 10.6&01.6 & 2 & 50.27&0.55 & +19.12&0.01 & 6.83&0.69 & 17.7& 3.4 & 10.68&0.15 \\
 33 & J121415.5+454003 & 12 14 18.6&0.6 & +45 40 09.7&04.7 & 2 & 6.14&0.21 & +4.95&0.06 & 0.93&0.10 & 12.9&14.6 & 6.01&0.40 \\
 34 & J121438.2+500646 & 12 14 40.4&0.7 & +50 06 55.2&07.2 & 1 & 2.41&0.14 & +31.69&0.08 & 0.15&0.02 & 25.5&10.8 & 2.76&0.14 \\
 35 & J121622.7+524422 & 12 16 23.9&0.8 & +52 44 18.3&07.3 & 3 & 4.41&0.23 & +21.58&0.06 & 7.07&1.00 &  \multicolumn{2}{c}{--} & \multicolumn{2}{c}{--} \\
 36 & J121839.5+502549 & 12 18 38.9&0.4 & +50 25 40.5&03.1 & 4 & 25.83&0.54 & +27.75&0.02 & 1.60&0.16 & 35.9& 9.5 & 7.26&0.36 \\
 37 & J121849.8+534620 & 12 18 50.6&0.4 & +53 46 16.4&04.4 & 3 & 4.10&0.10 & +13.49&0.02 & 3.04&0.34 &  \multicolumn{2}{c}{--} & \multicolumn{2}{c}{--} \\
 38 & J121916.1+552250 & 12 19 18.4&0.5 & +55 22 26.9&03.6 & 2 & 5.17&0.12 & +19.72&0.02 & 1.97&0.21 &  2.3& 8.6 & 11.48&0.47 \\
 39 & J121935.5+552828 & 12 19 33.6&0.7 & +55 28 20.9&05.7 & 2 & 4.76&0.21 & +22.00&0.06 & 0.60&0.07 & 15.9& 9.0 & 8.46&0.38 \\
 40 & J122106.2+454845 & 12 21 03.1&1.4 & +45 48 42.7&10.2 & 1 & 2.32&0.20 & +7.39&0.08 & 0.38&0.05 &  2.3&11.9 & 4.35&0.26 \\
 41 & J122156.7+454738 & 12 21 55.2&1.0 & +45 47 46.7&05.9 & 1 & 2.30&0.15 & +10.47&0.08 & 0.71&0.09 & 31.1&16.2 & 4.25&0.31 \\
 42 & J122607.8+473659 & 12 26 08.1&0.8 & +47 37 22.9&09.2 & 1 & 3.74&0.21 & +9.23&0.06 & 0.42&0.05 & 18.5& 5.3 & 6.58&0.18 \\
 43 & J123129.3+491539 & 12 31 28.2&0.9 & +49 15 35.9&08.5 & 1 & 1.46&0.09 & +14.48&0.06 & 0.82&0.10 &  \multicolumn{2}{c}{--} & \multicolumn{2}{c}{--} \\
 44 & J123234.7+482133 & 12 32 35.7&0.5 & +48 21 33.5&04.2 & 2 & 4.05&0.13 & +6.72&0.04 & 0.54&0.06 & 13.7& 5.2 & 5.21&0.14 \\
 45 & J123436.1+532225 & 12 34 34.5&0.4 & +53 22 44.2&03.2 & 4 & 6.03&0.16 & +7.38&0.03 & 2.30&0.24 &  \multicolumn{2}{c}{--} & \multicolumn{2}{c}{--} \\
 46 & J123506.6+562503 & 12 35 09.2&0.7 & +56 24 46.3&07.3 & 1 & 2.77&0.13 & +12.49&0.07 & 0.44&0.05 &  5.5&12.0 & 5.23&0.29 \\
 47 & J123527.8+531457 & 12 35 25.6&1.1 & +53 15 11.5&07.7 & 1 & 1.93&0.13 & +9.74&0.08 & 0.66&0.08 &  \multicolumn{2}{c}{--} & \multicolumn{2}{c}{--} \\
 48 & J123723.7+505717 & 12 37 22.8&0.5 & +50 57 14.8&04.2 & 3 & 3.63&0.11 & +13.85&0.03 & 3.87&0.45 &  \multicolumn{2}{c}{--} & \multicolumn{2}{c}{--} \\
 49 & J124007.1+533429 & 12 40 05.0&0.4 & +53 34 27.6&03.0 & 4 & 17.99&0.41 & +19.12&0.03 & 1.97&0.20 &  \multicolumn{2}{c}{--} & \multicolumn{2}{c}{--} \\
 50 & J124022.0+465638 & 12 40 20.5&0.3 & +46 56 54.7&03.2 & 2 & 9.68&0.20 & +8.59&0.02 & 1.65&0.17 & -4.1& 5.8 & 7.77&0.21 \\
 51 & J124115.3+514126 & 12 41 12.8&0.8 & +51 41 23.4&06.0 & 3 & 8.78&0.37 & +16.73&0.05 & 0.40&0.04 & 12.0& 3.9 & 7.66&0.15 \\
 52 & J124331.0+521941 & 12 43 30.3&0.9 & +52 19 42.6&07.0 & 3 & 3.68&0.19 & +13.70&0.05 & 2.89&0.34 & 15.9& 8.6 & 23.86&0.95 \\
 53 & J130145.1+540844 & 13 01 43.2&0.5 & +54 08 37.8&03.5 & 4 & 15.57&0.39 & +13.85&0.02 & 1.67&0.17 &  \multicolumn{2}{c}{--} & \multicolumn{2}{c}{--} \\
 54 & J130414.2+554136 & 13 04 11.7&4.8 & +55 42 04.6&40.3 & 1 & 3.11&0.83 & +19.40&2.08 & 0.20&0.06 & 49.8&15.4 & 3.91&0.28 \\
 55 & J130709.4+492140 & 13 07 10.6&0.9 & +49 21 41.7&08.3 & 2 & 2.52&0.13 & +14.58&0.07 & 2.47&0.30 &  \multicolumn{2}{c}{--} & \multicolumn{2}{c}{--} \\
 56 & J130748.6+471021 & 13 07 51.5&0.3 & +47 09 47.6&03.5 & 1 & 7.78&0.18 & +7.61&0.02 & 0.46&0.05 & 21.2& 4.4 & 6.96&0.14 \\
 57 & J130854.6+553047 & 13 08 56.7&1.5 & +55 30 58.0&10.0 & 2 & 2.08&0.19 & +15.47&0.09 & 0.79&0.11 &  \multicolumn{2}{c}{--} & \multicolumn{2}{c}{--} \\
 58 & J131634.4+493239 & 13 16 39.8&1.2 & +49 32 57.6&09.8 & 1 & 1.21&0.09 & +12.69&0.09 & 2.52&0.38 &  \multicolumn{2}{c}{--} & \multicolumn{2}{c}{--} \\
 59 & J132632.1+515413 & 13 26 32.4&1.8 & +51 54 16.2&15.1 & 3 & 2.68&0.27 & +20.40&0.11 & 0.37&0.05 &  5.9& 5.9 & 3.89&0.11 \\
 60 & J133437.2+563147 & 13 34 36.1&0.3 & +56 31 47.9&02.1 & 2 & 17.10&0.25 & +11.32&0.01 & 2.19&0.22 & 28.0& 8.1 & 3.93&0.14 \\
 61 & J133534.6+563114 & 13 35 34.0&0.5 & +56 31 09.8&03.5 & 2 & 9.02&0.26 & +7.79&0.04 & 1.15&0.12 & 13.3& 6.4 & 4.24&0.13 \\
 62 & J133922.6+464014 & 13 39 20.1&0.5 & +46 41 11.4&04.7 & 2 & 12.68&0.46 & +20.56&0.04 & 0.49&0.05 &  5.5& 7.3 & 7.70&0.25 \\
 63 & J134103.7+491532 & 13 41 04.5&0.4 & +49 15 26.4&05.2 & 1 & 7.18&0.18 & +9.73&0.02 & 4.24&0.47 &  \multicolumn{2}{c}{--} & \multicolumn{2}{c}{--} \\
 64 & J134545.3+533254 & 13 45 47.1&0.9 & +53 32 55.2&06.8 & 4 & 8.65&0.44 & +15.16&0.06 & 0.54&0.06 & 38.2&15.1 & 2.50&0.14 \\
 65 & J134548.2+564931 & 13 45 48.1&0.7 & +56 49 30.5&05.9 & 2 & 7.31&0.30 & +12.40&0.05 & 2.48&0.27 &  \multicolumn{2}{c}{--} & \multicolumn{2}{c}{--} \\
 66 & J134835.2+515605 & 13 48 35.5&0.9 & +51 56 04.7&05.7 & 1 & 2.27&0.10 & +15.82&0.05 & 1.41&0.16 &  \multicolumn{2}{c}{--} & \multicolumn{2}{c}{--} \\
 67 & J135140.2+564437 & 13 51 39.8&0.7 & +56 44 33.8&05.7 & 2 & 7.86&0.29 & +18.02&0.04 & 4.32&0.47 &  \multicolumn{2}{c}{--} & \multicolumn{2}{c}{--} \\
 68 & J135849.3+475503 & 13 58 49.3&0.7 & +47 54 51.1&06.6 & 3 & 4.73&0.18 & +18.99&0.05 & 1.63&0.18 &  \multicolumn{2}{c}{--} & \multicolumn{2}{c}{--} \\
 69 & J140227.2+520431 & 14 02 27.7&0.6 & +52 04 49.4&05.1 & 1 & 5.45&0.18 & +9.92&0.04 & 0.39&0.04 & 15.2& 4.1 & 7.69&0.14 \\
 70 & J140539.5+541137 & 14 05 39.0&0.5 & +54 11 40.7&04.1 & 3 & 8.06&0.25 & +14.82&0.03 & 4.43&0.48 &  \multicolumn{2}{c}{--} & \multicolumn{2}{c}{--} \\
 71 & J141946.0+542314 & 14 19 46.7&0.8 & +54 23 08.6&06.2 & 2 & 2.94&0.14 & +17.64&0.06 & 0.38&0.04 & 18.8& 3.6 & 2.22&0.03 \\
 72 & J142118.5+530328 & 14 21 18.9&1.0 & +53 03 19.7&10.3 & 2 & 3.22&0.23 & +18.87&0.09 & 0.26&0.03 &  \multicolumn{2}{c}{--} & \multicolumn{2}{c}{--} \\
 73 & J142308.4+505640 & 14 23 09.6&0.3 & +50 56 32.0&03.6 & 3 & 13.00&0.27 & +7.66&0.02 & 2.62&0.27 &  \multicolumn{2}{c}{--} & \multicolumn{2}{c}{--} \\
 74 & J143605.2+534920 & 14 36 03.7&0.5 & +53 48 53.2&05.8 & 1 & 2.04&0.07 & +12.42&0.04 & 0.52&0.06 &  0.4&13.7 & 5.22&0.33 \\
 75 & J143642.3+560816 & 14 36 40.5&1.1 & +56 07 58.4&09.7 & 1 & 1.61&0.09 & +15.45&0.05 & 0.51&0.06 &  \multicolumn{2}{c}{--} & \multicolumn{2}{c}{--} \\
 76 & J143912.1+541829 & 14 39 14.4&0.6 & +54 18 21.9&04.4 & 1 & 2.12&0.06 & +11.63&0.03 & 1.09&0.12 &  \multicolumn{2}{c}{--} & \multicolumn{2}{c}{--} \\
 77 & J144248.0+535416 & 14 42 52.4&0.7 & +53 54 20.2&04.9 & 1 & 3.57&0.12 & +8.26&0.03 & 0.26&0.03 &  9.3& 3.7 & 10.48&0.19 \\
 78 & J144301.5+520136 & 14 43 01.3&0.3 & +52 01 37.9&03.1 & 4 & 98.01&2.24 & +15.06&0.02 & 0.63&0.06 & 13.3& 0.5 & 5.48&0.01 \\
 79 & J144542.5+474919 & 14 45 41.7&0.8 & +47 49 00.2&08.6 & 3 & 5.97&0.37 & +16.42&0.07 & 0.44&0.05 & 14.4& 2.9 & 13.40&0.18 \\
 80 & J145012.6+471046 & 14 50 11.9&0.3 & +47 10 46.5&03.0 & 2 & 6.21&0.13 & +6.81&0.02 & 2.23&0.23 & -1.5& 7.7 & 8.58&0.31 \\
 81 & J145046.1+530005 & 14 50 45.4&0.5 & +52 59 59.6&05.1 & 4 & 7.53&0.26 & +16.92&0.05 & 1.08&0.12 &  \multicolumn{2}{c}{--} & \multicolumn{2}{c}{--} \\
 82 & J145356.5+502731 & 14 53 56.1&0.6 & +50 27 26.5&05.6 & 1 & 2.61&0.10 & +14.07&0.04 & 1.16&0.13 &  \multicolumn{2}{c}{--} & \multicolumn{2}{c}{--} \\
 83 & J145426.6+514544 & 14 54 26.8&0.7 & +51 45 53.6&06.5 & 2 & 3.58&0.15 & +17.81&0.05 & 1.88&0.22 & 15.4&11.7 & 7.47&0.40 \\
 84 & J145427.4+512436 & 14 54 27.6&1.0 & +51 24 44.3&09.6 & 2 & 2.40&0.17 & +20.08&0.08 & 0.38&0.05 & 17.5& 7.9 & 4.16&0.15 \\
 85 & J145854.8+464917 & 14 58 57.5&0.3 & +46 49 20.2&03.2 & 2 & 7.49&0.16 & -5.01&0.02 & 1.61&0.17 &  \multicolumn{2}{c}{--} & \multicolumn{2}{c}{--} \\
 86 & J150013.4+501550 & 15 00 13.4&0.8 & +50 15 43.6&06.7 & 2 & 2.46&0.12 & +13.08&0.06 & 0.70&0.08 & 29.4&11.1 & 5.95&0.33 \\
 87 & J150017.0+543605 & 15 00 19.0&1.6 & +54 36 03.9&13.4 & 1 & 1.42&0.15 & +17.84&0.11 & 1.09&0.17 &  \multicolumn{2}{c}{--} & \multicolumn{2}{c}{--} \\
 88 & J150048.7+475113 & 15 00 48.8&0.5 & +47 51 11.9&05.0 & 4 & 6.42&0.21 & +6.18&0.04 & 1.09&0.12 &  2.6& 1.5 & 8.05&0.06 \\
 89 & J150439.0+503005 & 15 04 42.9&1.1 & +50 30 03.0&06.8 & 1 & 2.70&0.15 & +6.05&0.08 & 0.62&0.07 & 10.3& 8.7 & 18.00&0.67 \\
 90 & J150609.8+513531 & 15 06 08.7&0.5 & +51 35 18.2&04.4 & 1 & 3.30&0.09 & +9.41&0.03 & 3.13&0.37 &  \multicolumn{2}{c}{--} & \multicolumn{2}{c}{--} \\
 91 & J150644.2+493355 & 15 06 43.7&0.5 & +49 33 51.9&03.7 & 1 & 2.65&0.07 & +8.90&0.03 & 2.81&0.34 & -51.8&15.1 & 3.62&0.25 \\
$^{(*)}$ & J111538.5+503024 & 11 15 37.5&0.2 & +50 30 09.7&02.1 & 2 & 5.83&0.07 & +2.69&0.01 & 2.95&0.30 & \multicolumn{2}{c}{--} & \multicolumn{2}{c}{--} \\
\end{longtable}
\end{landscape}
}

\longtab[2]{
\begin{longtable}{lllll}
  \caption{Radio and optical descriptions of polarized sources}\\
  \label{table:ids}\\
    \hline
    ID&Source type&LOFAR?&Optical ID?&Comments\\
    \hline
    \endhead
    \hline
    \endfoot
1	&	FR~{\sc ii} hotspot	&	N	&	Y	&	Large radio galaxy\\
2	&	Compact	&	Y	&	Y\\		
3	&	FR~{\sc ii} hotspot	&	N	&	Y	&	Large radio galaxy\\
4	&	FR~{\sc ii} hotspot	&	Y	&	Y &\\		
5	&	FR~{\sc ii} hotspot	&	Y	&	Y &\\		
6	&	FR~{\sc ii} hotspot	&	Y	&	Y	&	Large radio galaxy\\
7	&	FR~{\sc ii} hotspot	&	Y	&	Y &\\		
8	&	Compact	&	Y	&	Y\\		
9	&	FR~{\sc ii} hotspot	&	Y	&	Y &\\		
10	&	FR~{\sc ii} hotspot	&	Y	&	Y & \\		
11	&	FR~{\sc ii} hotspot	&	Y	&	Y &\\		
12	&	FR~{\sc ii} hotspot	&	Y	&	Y &\\	
13	&	FR~{\sc ii} hotspot	&	Y	&	Y&\\		
14	&	FR~{\sc ii} hotspot	&	Y	&	Y	&	Large radio galaxy\\
15	&	Compact	&	Y	&	Y&\\		
16	&	FR~{\sc ii} hotspot	&	N	&	Y&\\		
17	&	Compact double	&	Y	&	Y&\\		
18	&	FR~{\sc ii} hotspot	&	N	&	Y&\\		
19	&	FR~{\sc ii} hotspot	&	Y	&	N	&	Foreground star hinders ID\\
20	&	Compact+halo	&	Y	&	Y&\\		
21	&	Extended, unclear	&	Y	&	Y&\\		
22	&	Compact double	&	Y	&	N&\\		
23	&	Compact double	&	Y	&	Y&\\		
24	&	FR~{\sc ii} hotspot	&	Y	&	Y&\\		
25	&	FR~{\sc ii} hotspot	&	Y	&	Y	&	Same as \#24\\
26	&	FR~{\sc ii} hotspot	&	N	&	Y&\\		
27	&	FR~{\sc ii} hotspot	&	Y	&	Y&\\		
28	&	Extended, unclear	&	Y	&	Y&\\		
29	&	FR~{\sc ii} hotspot	&	Y	&	Y&\\		
30	&	FR~{\sc ii} hotspot	&	Y	&	Y&\\		
31	&	Compact	&	Y	&	N&\\		
32	&	FR~{\sc ii} hotspot	&	Y	&	Y&\\		
33	&	FR~{\sc ii} hotspot	&	N	&	Y	&	{\it WISE} ID only\\
34	&	FR~{\sc ii} hotspot	&	Y	&	N	&	{\it WISE} ID only\\
35	&	FR~{\sc ii} hotspot	&	Y	&	Y&\\		
36	&	FR~{\sc ii} hotspot	&	Y	&	Y	&	Large radio galaxy\\
37	&	Compact double	&	Y	&	Y	&	{\it WISE} ID only\\
38	&	FR~{\sc ii} hotspot	&	Y	&	Y	&	{\it WISE} ID only\\
39	&	FR~{\sc ii} hotspot	&	Y	&	Y	&	{\it WISE} ID only\\
40	&	FR~{\sc ii} hotspot	&	N	&	Y&\\		
41	&	FR~{\sc ii} hotspot	&	N	&	Y	&	Structure unclear\\
42	&	FR~{\sc ii} hotspot	&	Y	&	Y&\\		
43	&	FR~{\sc i} jet/lobe	&	Y	&	Y&\\		
44	&	Compact	&	Y	&	Y&\\		
45	&	FR~{\sc ii} hotspot	&	Y	&	Y	&	Large radio galaxy (N lobe of \#47)\\
46	&	FR~{\sc ii} hotspot	&	N	&	Y&\\		
47	&	FR~{\sc ii} hotspot	&	Y	&	Y	&	Large radio galaxy (S lobe of \#45)\\
48	&	Compact	&	Y	&	Y&\\		
49	&	FR~{\sc ii} hotspot	&	Y	&	Y&\\		
50	&	FR~{\sc ii} hotspot	&	Y	&	Y&\\		
51	&	FR~{\sc ii} hotspot	&	Y	&	Y&\\		
52	&	FR~{\sc ii} hotspot	&	Y	&	Y?	&	ID uncertain\\
53	&	FR~{\sc ii} hotspot	&	Y	&	Y&\\		
54	&	FR~{\sc ii} hotspot	&	Y	&	Y&\\		
55	&	Compact	&	Y	&	Y&\\		
56	&	FR~{\sc ii} hotspot	&	Y	&	Y&\\		
57	&	FR~{\sc ii} hotspot	&	Y	&	Y&\\		
58	&	FR~{\sc ii}	&	Y	&	Y	&	Position is $\sim 1$ arcmin from core\\
59	&	Compact	&	Y	&	Y&\\		
60	&	FR~{\sc ii} hotspot	&	Y	&	Y&\\		
61	&	Compact	&	Y	&	N&\\		
62	&	FR~{\sc ii} hotspot	&	Y	&	Y&\\		
63	&	FR~{\sc ii} hotspot	&	Y	&	Y&\\		
64	&	FR~{\sc ii} hotspot or jet knot	&	Y	&	Y	&	Complex source\\
65	&	Compact double	&	Y	&	Y&\\		
66	&	FR~{\sc ii} hotspot	&	Y	&	Y	&	{\it WISE} ID only\\
67	&	FR~{\sc ii} hotspot	&	Y	&	Y	&	{\it WISE} ID only\\
68	&	FR~{\sc ii} hotspot	&	Y	&	Y&\\		
69	&	FR~{\sc ii} hotspot	&	Y	&	Y&\\		
70	&	FR~{\sc ii} hotspot	&	Y	&	Y&\\		
71	&	Compact	&	Y	&	Y&\\		
72	&	Compact double	&	Y	&	Y&\\		
73	&	FR~{\sc ii} hotspot	&	N	&	Y?	&	ID uncertain\\
74	&	Compact double	&	Y	&	Y&\\		
75	&	FR~{\sc ii} hotspot	&	Y	&	Y&\\		
76	&	FR~{\sc ii} hotspot	&	Y	&	Y&\\		
77	&	FR~{\sc ii}	&	Y	&	Y&\\		
78	&	FR~{\sc ii} hotspot	&	Y	&	Y&\\		
79	&	FR~{\sc ii} hotspot	&	Y	&	Y&\\		
80	&	Compact	&	Y	&	Y&\\		
81	&	FR~{\sc ii} hotspot	&	Y	&	Y&\\		
82	&	Compact	&	Y	&	Y&\\		
83	&	FR~{\sc ii} hotspot	&	Y	&	Y&\\		
84	&	Compact/jet	&	Y	&	Y&\\		
85	&	FR~{\sc ii} hotspot	&	Y	&	Y	&	{\it WISE} ID only\\
86	&	FR~{\sc ii}	&	Y	&	Y &\\		
87	&	Compact	&	Y	&	Y&\\		
88	&	Compact	&	Y	&	Y&\\		
89	&	FR~{\sc ii}	&	Y	&	Y&\\	
90	&	FR~{\sc ii} hotspot	&	Y	&	Y	&	Large radio galaxy\\
91	&	Compact	&	Y	&	Y&\\	
\end{longtable}
}

\end{document}